\documentclass[journal=jctcce,manuscript=article,layout=traditional]{achemso}

\usepackage{graphicx} % Required for inserting images
\usepackage{amsmath}
\usepackage{algorithm,algpseudocode}
\usepackage{amsfonts}
\usepackage{amssymb,bm}
\usepackage{tikz}
\usepackage{cleveref}
\usepackage{subfiles}
\usepackage{booktabs}

\usepackage{siunitx}

\usepackage[nice]{nicefrac}

\usepackage{dsfont}
\usepackage{esvect}
\usepackage{url} 

\DeclareSIUnit\angstrom{\text {Å}}

\DeclareMathOperator{\Tr}{Tr}
\let\oldforall\forall
\renewcommand{\forall}{\, \oldforall \, }

\let\oldexist\exists
\renewcommand{\exists}{\oldexist \: }

\usetikzlibrary{positioning}
\usetikzlibrary{tikzmark}
\usetikzlibrary{calc}
\usetikzlibrary{backgrounds}

\DeclareUnicodeCharacter{03C3}{\(\sigma\)}

\tikzset{%
  add/.style args={#1 and #2}{to path={%
 ($(\tikztostart)!-#1!(\tikztotarget)$)--($(\tikztotarget)!-#2!(\tikztostart)$)%
  \tikztonodes}}
} 

\title{Periodic implementation of the random phase approximation with numerical atomic orbitals and dual reciprocal space grids}

\author{Edoardo Spadetto}
\email{e.spadetto@vu.nl}
\affiliation{Theoretical Chemistry, Vrije Universiteit, De Boelelaan 1108, 1081 HZ Amsterdam, The Netherlands}
\alsoaffiliation{Software for Chemistry and Materials NV, NL, 1081HV, Amsterdam, The Netherlands}

\author{Pier Herman Theodoor Philipsen}
\email{philipsen@scm.com}
\affiliation{Software for Chemistry and Materials NV, NL, 1081HV, Amsterdam, The Netherlands}

\author{Arno Förster}
\email{a.t.l.foerster@vu.nl}
\affiliation{Theoretical Chemistry, Vrije Universiteit, De Boelelaan 1108, 1081 HZ Amsterdam, The Netherlands}

\author{Lucas Visscher}
\email{l.visscher@vu.nl}
\affiliation{Theoretical Chemistry, Vrije Universiteit, De Boelelaan 1108, 1081 HZ Amsterdam, The Netherlands}


\DeclareUnicodeCharacter{2212}{-}

\date{\today}

\begin{document}

\begin{abstract}
The random phase approximation (RPA) has emerged as a prominent first-principles method in material science, particularly to study the adsorption and chemisorption of small molecules on surfaces. However, its widespread application is hampered by its relatively high computational cost.
Here, we present a well-parallelised implementation of the RPA with localised atomic orbitals and pair-atomic density fitting, which is especially suitable for studying two-dimensional systems. Through a dual $\bm{k}$-grid scheme, we achieve fast and reliable convergence of RPA correlation energies to the thermodynamic limit. We demonstrate the efficacy of our implementation through an application to the adsorption of CO on MgO(001) using PBE input orbitals (RPA@PBE). Our calculated adsorption energy is in excellent agreement with previously published RPA@PBE studies, but, as expected, overestimates the experimentally available adsorption energies as well as recent CCSD(T) results. 
\end{abstract}

\begin{tocentry}
\centering
\vfill
\includegraphics[scale=0.03]{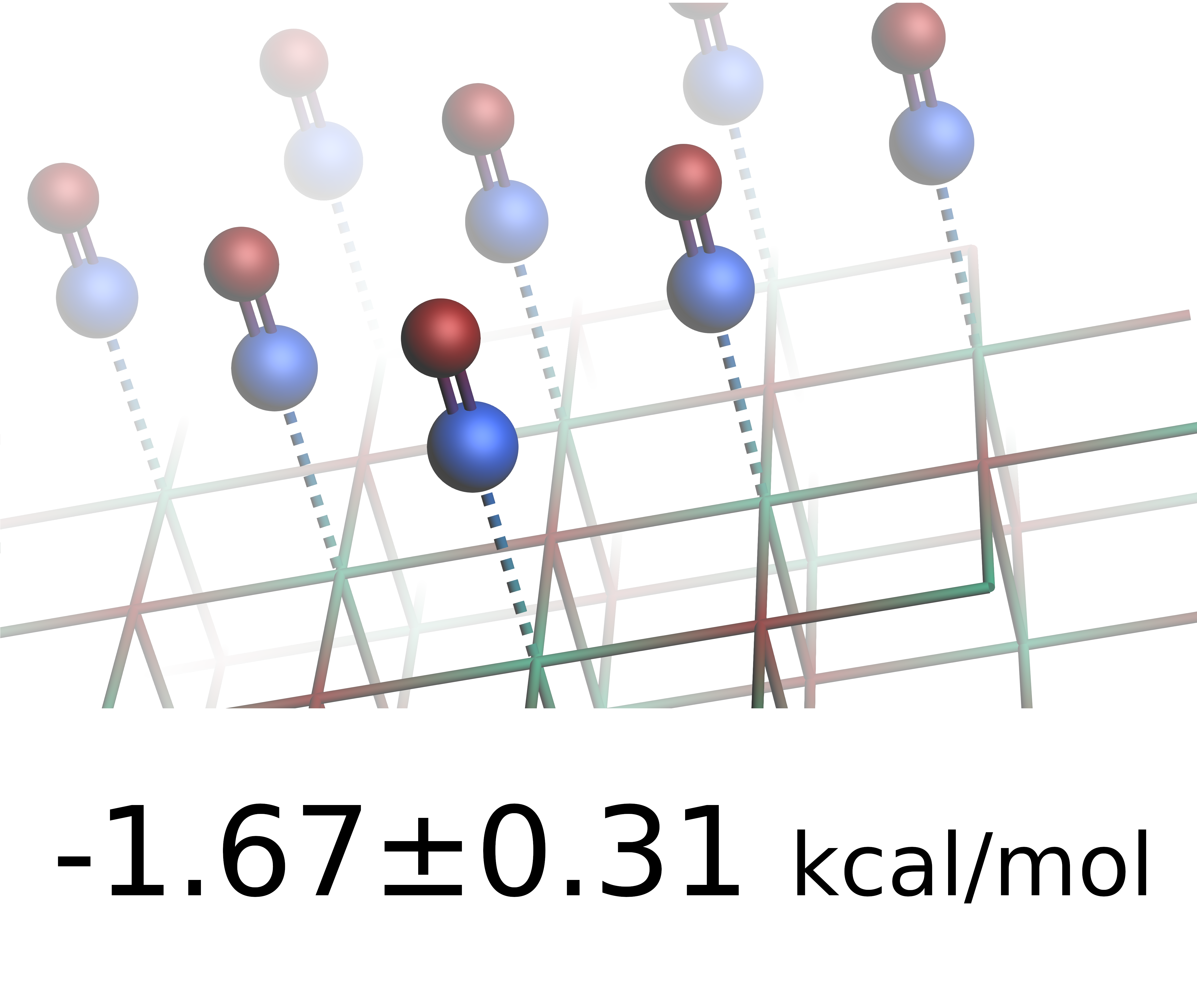}
\vfill
\end{tocentry}

\maketitle

% \tableofcontents
% \input{outline}

\section{\label{sec::introduction}Introduction}
The Random Phase Approximation (RPA) is a well-established method in quantum chemistry and solid-state physics.\cite{Ren2012a} Originally developed by Bohm and Pines while working on plasmonic oscillation in the jellium model,\cite{bohmpines1,bohmpines2,Bohm1953} it is equivalent to approximating the correlation energy through a summation of ring diagrams within diagrammatic perturbation theory.\cite{Hubbard1957, Gell-Mann1957, Mattuck1992} Equivalently, it can be derived from the Klein functional\cite{Klein1961} using a non-interacting single particle Green function\cite{kleinF,Klein1961,caruso} or within the Adiabatic-Connection (AC) Fluctuation Dissipation theorem (ACFDT).\cite{Langreth1975, Langreth1977} Ring diagrams describe one of the most important signatures of electron correlation,\cite{VanLoon2021} the screening of electron-electron interactions at large interelectronic distances. Therefore, the RPA is especially accurate for reactions dominated by the difference in the long-range correlation energies of the reactants, as caused, for instance, by non-covalent interactions.\cite{Harl2008, Kresse2009, Lebegue2010} 

The RPA exchange-correlation (xc) energy is fully non-local and depends on virtual orbitals, placing itself at the top of Perdew's \textit{Jacob's ladder} \cite{Perdew2001} of density functional theory (DFT).\cite{Hohenberg1964, Kohn1965} Therefore, it is computationally more involved than generalized gradient approximations (GGA).
While density functionals including empirical dispersion corrections\cite{Tkatchenko2009a, Stefan2011} are cheaper options for treating non-covalent interactions and give accurate results in many scenarios, they may break down in highly anisotropic or highly polarizable systems.\cite{Dobson_2012} Non-local dispersion corrections are more advanced alternatives but are also computationally more involved.\cite{Vydrov2010}

Compared to wave-function-based methods, the RPA includes interactions present both in second-order Møller--Plesset perturbation theory (MP2) and Coupled Cluster (CC) theory. \cite{Coester1960,cc1,Cizek1966,Stanton1993} The development of CC implementations for periodic systems is a relatively recent development\cite{Zhang2019, Schafer2021, Gruber2018, McClain2017, Vo2024, Ye2024a} and applications of these methods to the adsorption of CO on the MgO(001) surface\cite{Ye2024, Shi2023a} demonstrate their immense potential in applications to molecule-surface interactions. 
As shown by Scuseria, Henderson, and coworkers,\cite{Scuseria2008a, Scuseria2013} RPA can be seen as a simplification of coupled cluster with single and double excitations (CCSD), including fewer interaction channels but having the advantage of much lower computational cost.

RPA and MP2 both share the second-order ring diagram, but this term diverges in strongly polarizable\cite{Nguyen2020} small band gap systems or metals in the thermodynamic limit.\cite{Gruneis2010a} While MP2 is increasingly applied to solids\cite{Gruneis2010a, DelBen2012, Schafer2017} and these shortcomings might be alleviated using regularization techniques,\cite{Keller2022} the RPA is widely applicable without such corrections.\cite{Gell-Mann1957,Gruneis2010a,Mattuck1992} To combine the benefits of MP2 and RPA, the addition of second-order screened exchange (SOX) corrections has also been explored extensively for molecules\cite{Paier2012,  Ren2013, Bates2013, Forster2022a} and homogeneous electron gases.\cite{Gruneis2009, Hummel2019} SOX corrections come however with increased computational cost and tend to deteriorate the good description of stretched bonds within the RPA,\cite{Henderson2010, Paier2012, Ren2013, Bates2013} which is important to describe transition states.

The many beneficial characteristics of RPA are key factors behind the increasing popularity of the RPA for applications in heterogeneous catalysis. Here, the rate-determining step is the chemisorption (involving an effective chemical bond between the surface and the reactant) of a reactant (typically small molecules) followed by their dissociation into reactive fragments.\cite{Kroes2021} Modelling such molecule-surface interactions accurately requires a reliable description of long-range dispersion forces and reaction barrier heights.\cite{Kroes2021} In many instances, GGAs are not suitable to study these processes,\cite{Gerrits2020, Kroes2021} and the RPA promises much higher accuracy.\cite{Schmidt2018, Oudot2024} For this reason the RPA has been applied to model a wide array of molecule-surface interactions\cite{Rohlfing2008, Ren2009, Schimka2010, GarridoTorres2017, Sheldon2021, Wei2021, Liu2023a, Oudot2024, Weinberg2024, Clary2025} and is predicted to play an increasingly prominent role in computational catalysis in the future.\cite{Sauer2024, Kroes2025} More advanced ACFDT methods like adiabatic xc kernel methods\cite{Olsen2012a, Olsen2014, Olsen2019} or $\sigma$-functionals\cite{Trushin2021, Fauser2021, Erhard2022} can boost the accuracy of the RPA at similar computational cost and their implementations can be realized through minor modifications of the underlying RPA algorithms. We focus here on the efficient implementation of the RPA, which is an important stepping stone to these more advanced methods.

While being significantly cheaper than wave-function-based methods, RPA calculations are computationally much more demanding than conventional DFT. In their canonical formulations\cite{ren2012, Ren2021} RPA calculations scale as $N^4_AN^2_{\bm{k}}$ with $N_A$ being the number of atoms in the unit cell and $N_{\bm{k}}$ being the number of $\bm{k}$-points.\cite{Harl2008} Applications of the RPA and other post-SCF methods to heterogenous catalysis therefore often rely on embedding approaches\cite{Boese2013, Alessio2019, Yin2018, Wei2023, Gujarati2023, Sheldon2024, Huang2025} and further algorithmic developments are needed to make full RPA calculations routine for molecule-surface interactions. Algorithms based on the space-time formulation\cite{H.N.Rojas1995, Rieger1999} of the $GW$ approximations\cite{Hedin1965} lower the scaling of RPA calculations to formally $N^3_AN_{\bm{k}}$.\cite{Kaltak2014, Kaltak2014a, Kutepov2020, Yeh2023, Yeh2024a, Graml2024, Shi2024, Shi2025} In practice, the time-determining step of these calculations is the calculation of the RPA polarizability, which can be done with sub-quadratic scaling, and therefore the scaling is sub-quadratic for practical system sizes.\cite{Graml2024, Shi2024, Shi2025} These algorithms come with a higher prefactor than canonical implementations, and for this reason, they are useful for large $\bm{k}$-grids and large unit cells. Important potential use-cases include Moiré superlattices,\cite{Graml2024} point-defects, or molecule-surface interactions with low coverages.\cite{Kelsall2021}

Many molecule-surface interactions can however be modelled with relatively small unit cells for which canonical RPA implementations are more suitable. Here, it is important to account for the reduced 2-dimensional (2D) periodicity of the system. In plane-wave-based implementations, the 2D geometry is replicated periodically along the non-periodic spatial direction, and large replica separations are required to prevent unphysical interactions between them.\cite{pwdipoles, DecoupleImages, modCoulombforimages, yambo,gianozzi09}. The number of plane waves required to reach a given energy cut-off increases significantly with the lattice constant, making such calculations computationally demanding.\cite{Engel2014} To address this, plane-wave simulations of molecule–surface interactions often employ truncated Coulomb potentials.\cite{Rozzi2006, Castro2009} Despite these techniques, relatively large replica separations of 10–20~\si{\angstrom} are still necessary.\cite{Guandalini2023} This approach also introduces challenges in Brillouin zone integration, necessitating even denser $\bm{k}$-point grids.\cite{Huser2013, Qiu2016}

For these low-dimensional systems, atomic orbital-based implementations are more suitable since they can naturally describe the non-periodic behavior of the wave function.\cite{Susi19} The most commonly employed localized basis sets for post-self-consistent field (SCF) calculations in periodic systems are Gaussian-type orbitals (GTO)\cite{dovesi20, Boys1950, mcmurchies78, obara86,gill91} as implemented for instance in 
% CRYSTAL,\cite{Pisani1980, Dovesi1983, Dovesi2018, Dovesi2020} 
pySCF\cite{McClain2017, Sun2017, Sun2020, Ye2022, Vo2024} or CP2K\cite{DelBen2012, DelBen2015, Wilhelm2017, Kuhne2020a, Stein2024} and numerical atomic orbitals (NAO) as implemented in FHI-AIMS\cite{Blum2009,Ren2021, Qu2022, Shi2024} or ABACUS.\cite{Lin2020, Lin2021, Li2016b, Lin2023} NAO basis sets are both compact and highly flexible and can be used to numerically represent GTOs, Slater-type orbitals (STO), or molecular DFT orbitals. Choosing STOs as their functional form, the exponential decay of the wave function far from a finite system can be better represented.\cite{Zhang2013} This property is important when modeling 2D systems, and is difficult to achieve with plane waves.\cite{Engel2014} In practice, this enables energies closer to the infinite basis set limit for the same number of basis functions.\cite{Spadetto2023}

This work reports a novel NAO-based RPA implementation in the BAND\cite{TeVelde1991} module of the Amsterdam modeling suite (AMS).\cite{Baerends2025} Our implementation relies on the pair-atomic density fitting (PADF) approximation\cite{Forster2020, Spadetto2023} (also known as local RI\cite{Lin2020, Lin2021, Shi2024, Shi2025} or concentric atomic density fitting\cite{Hollman2014, Hollman2017, Wang2020a}) which expands products of atomic basis functions in an additional fit set of functions which are centered only on the same atoms from which a specific basis set product originates.\cite{Baerends1973,Forster2020,Watson2003,Manzer2015} PADF has been used extensively within the space-time method to realise low-scaling implementations of RPA and $GW$ for finite\cite{Forster2020,Forster2020b} and periodic systems,\cite{Shi2024, Shi2025} and for periodic Hartree-Fock.\cite{Lin2020, Lin2021, Lin2023} Here, we use it only to reduce the size of involved tensors and consequently memory demands. 

Inspired by the HF implementation of \citet{Irmler2018}, we introduce a scheme to dampen the Coulomb potential at long distances depending on the size of the employed $\bm{k}$-grid. Using mono- and bilayer hexagonal boron nitride (h-BN) as well as the adsorption of CO on monolayer MgO(001) as test systems, we demonstrate this method to be numerically stable and leading to a rapid convergence to the thermodynamic limit in combination with a dual $\bm{k}$-grid scheme (inspired by the staggered-mesh method\cite{Xing2021, Xing2022}) which we introduce here as well.  We also describe a parallelisation scheme, which handles reciprocal space and frequency grid integration, achieving near-perfect parallel efficiency for thousands of cores on multiple nodes. Finally, as a practical application for our implementation, we calculate the adsorption energy of CO on the MgO(001) surface and demonstrate good agreement of our result to previous periodic RPA calculations.\cite{Bajdich2015} 

 \section{\label{sec::theory}Theory}
 \subsection{RPA correlation energy from Adiabatic connection}
From the ACFDT, the RPA Correlation energy can be written as\cite{Langreth1975, Ren2012a} 
 \begin{align}
 E^{RPA}_{corr}
 & = - \frac{1}{2\pi} \int_{0}^{\infty} \mathrm{d} \omega \text{Tr} \Bigg[ \sum_{n=1}  \frac{1}{n} [ Z(\bm{r},\bm{r}', i\omega) ] ^{n}   - Z(\bm{r},\bm{r}',i\omega) \Bigg ] \;,
 \end{align}
 with
 \begin{equation}
     Z(\bm{r},\bm{r}',i\omega) = \int \mathrm{d} r'' P^{(0)}(\bm{r},\bm{r}'',i\omega)v(\bm{r}'',\bm{r}') \;,
 \end{equation}
 and the trace operator for a generic two variable function $A(\bm{r},\bm{r}')$  defined as
 \begin{equation}
     \text{Tr}[A(\bm{r},\bm{r}')]  = \int \mathrm{d} r \mathrm{d} r' A(\bm{r},\bm{r}') \delta (\bm{r}-\bm{r}') \; .
 \end{equation}
Here,  $P^{(0)}(\bm{r},\bm{r}', i\omega)$ is the irreducible  RPA polarizability, and $v(\bm{r},\bm{r}')=\frac{1}{|\bm{r}-\bm{r}'|}$ is the Coulomb potential.  By following the derivation given in the appendix, we can expand this equation in a basis $F = \left\{f(\bm{r},\bm{q})\right\}$ with $\bm{q}$ being differences between points in the first Brillouin zone (1BZ). We obtain
\begin{align}
\label{working-equation}
    E^{RPA}_{corr} 
    = - \frac{1}{2\pi} \sum_{\bm{q} \in \text{1BZ}} \int_{0}^{\infty} \mathrm{d} \omega \Bigg[ \log \left\{ \det \left[\mathbf{1}-\mathbf{Z}(\bm{q},i\omega) \right] \right\} - \text{Tr} \left [ \mathbf{Z}(\bm{q},i\omega) \right ] \Bigg ] \;,
\end{align}
where $\mathbf{Z} = \mathbf{v}^{\nicefrac{1}{2}} \mathbf{P}^{(0)}  \mathbf{v}^{\nicefrac{1}{2}}$, and the matrix elements of $\mathbf{P}^{(0)}$ are
\begin{align}
P^{(0)}_{\alpha\beta}(\bm{q},i\omega) = 
\langle f_{\alpha}(\bm{r},\bm{q}) |P^{(0)}(\bm{r},\bm{r}',i\omega) | f_{\beta}(\bm{r}',\bm{q}) \rangle \;.
    \label{eq:polExpansion}
\end{align}
The matrix $\mathbf{v}^{\nicefrac{1}{2}}(\bm{q})$ is the square root of the Coulomb potential with matrix elements
\begin{align}
    v_{\alpha \beta}(\bm{q}) = \langle f_{\alpha}(\bm{r},\bm{q}) | \frac{1}{|\bm{r}-\bm{r}'|}  | f_{\beta}(\bm{r}',\bm{q}) \rangle \;.
    \label{eq:vpot}
\end{align}
In general, for both  \ref{eq:polExpansion} and \ref{eq:vpot}, matrix elements between fit functions from different $\bm{q}$ and $\bm{q}'$  points can be defined. 
However these off-diagonal elements are zero, making $P^{(0)}_{\alpha\beta}(\bm{q},\omega)$ and $v_{\alpha\beta}(\bm{q})$ diagonal in $\bm{q}$. We therefore have simplified the notation and use only one $\bm{q}$ value to index these matrix elements.

\subsection{Bloch summation functions RI}
After having summarized the key relations to obtain $E^{RPA}_{corr}$, we will show in the following how the matrix elements of $P^{(0)}$ can be calculated. We will refer to the basis $F$ as fit set, to distinguish it from the primary basis set that is used to expand the orbitals in and which is labelled as $X$. We define $X = \{\chi_{\mu} (\bm{r},\bm{k}_{n})\} $ where the $\bm{k}_{n}$ vector defines the basis functions at its specific reciprocal point in the first Brillouin zone. The set of all $\bm{k}$-points will be called $K_{G}$.
Similarly, the fit set $F$ includes functions  $\{f_{\alpha} (\bm{r},\bm{q}_{n})\}$ which depend on all the reciprocal vectors $\bm{q}_{n} \equiv  \bm{k}_{m}-\bm{k}_{l} $, obtained as differences between all pairs $\bm{k}_{m}$,$\bm{k}_{l} \in K_{G}$. The set of all $\bm{q}$-vectors defines another grid $Q_{G}$.
Both types of basis functions are related to their real-space analogues via Fourier transforms,
\begin{equation}
    \chi_{\mu} (\bm{r},\bm{k}) = \sum_{\bm{R}} \chi_{\mu} (\bm{r}-\bm{R}) e^{i\bm{k}\cdot\bm{R}} 
    \label{eq:blochsumdefs}
\end{equation}
and 
\begin{equation}
    \label{eq:blochsumdefs2}
        f_{\alpha}(\bm{r},\bm{q}) = \sum_{\bm{R}}  f_{\alpha} (\bm{r}-\bm{R})e^{i\bm{q}\cdot\bm{R}} \;,
\end{equation}
where the vectors $\bm{R}$ enumerate the unit cells.
The fit set is related to the primary basis through the PADF equations\cite{Spadetto2023} (also known as local RI\cite{Lin2020, Lin2021, Shi2024, Shi2025}) as
\begin{align}
    \chi_{\mu \in A}(\bm{r}-\bm{R})\chi_{\nu \in B}(\bm{r}-\bm{R}') \approx \sum_{\alpha \in A} C^{\bm{R}',\bm{R}}_{\nu\mu\alpha} f_{\alpha}(\bm{r}-\bm{R})  +\sum_{\beta \in B} C^{\bm{R},\bm{R}'}_{\mu\nu\beta} f_{\beta}(\bm{r}-\bm{R}')  \;.
\end{align}
The fit functions on the \emph{r.h.s.} are thus centred on the same two atoms with indices $A$ and $B$ as the two primary basis functions. Utilising this definition, we next consider this product in reciprocal space
\begin{equation}
\begin{aligned}
   \chi^{*}_{\mu}(\bm{r},\bm{k})\chi_{\nu}(\bm{r},\bm{k}+\bm{q}) & = \sum_{\bm{R}} \chi_{\mu}(\bm{r}-\bm{R}) e^{-i\bm{k}\cdot\bm{R}} \sum_{\bm{R}'} \chi_{\nu}(\bm{r}-\bm{R}')e^{i(\bm{k}+\bm{q})\cdot\bm{R}'} \\ 
    &= \sum_{\bm{R}\bm{R}'} e^{i(\bm{k}+\bm{q})\cdot\bm{R}'}e^{-i\bm{k}\cdot\bm{R}}  \chi_{\mu}(\bm{r}-\bm{R}) \chi_{\nu}(\bm{r}-\bm{R}') \\ 
    &\approx \sum_{\bm{R}\bm{R}'} e^{i(\bm{k}+\bm{q})\cdot\bm{R}'}e^{-i\bm{k}\cdot\bm{R}}  \left[ \sum_{\alpha \in A} C_{\nu \mu \alpha}^{\bm{R}',\bm{R}} f_{\alpha} (\bm{r}-\bm{R}) + \sum_{\beta \in B} C_{\mu \nu \beta}^{\bm{R},\bm{R}'} f_{\beta}(\bm{r}-\bm{R}') \right] \;.
\end{aligned}
\end{equation}
By grouping the phase factors on both the atomic summations and by employing translational symmetry of fit coefficients, we obtain
\begin{equation}
\begin{aligned}
   \chi^{*}_{\mu}(\bm{r},\bm{k})\chi_{\nu}(r,\bm{k}+\bm{q}) &\approx \sum_{\alpha \in A} \sum_{\bm{R}\bm{R}'} e^{i(\bm{k}+\bm{q})\cdot\bm{R}'}e^{-i\bm{k}\cdot \bm{R}}  C_{\nu \mu \alpha}^{\bm{R}',\bm{R}} f_{\alpha} (\bm{r}-\bm{R}) + \\ &   \sum_{\beta \in B} \sum_{\bm{R}\bm{R}'} e^{i(\bm{k}+\bm{q})\bm{R}'}e^{-i\bm{k}\cdot\bm{R}} C_{\mu \nu \beta}^{\bm{R},\bm{R}'} f_{\beta}(r-\bm{R}')  \\
  & =\sum_{\alpha \in A} \sum_{\bm{R'}} e^{i(\bm{k}+\bm{q})\cdot(\bm{R}-\bm{R}')}  C_{\nu \mu \alpha}^{0,(\bm{R}'-\bm{R})}\sum_{R} e^{i\bm{q}\cdot\bm{R}}  f_{\alpha} (\bm{r}-\bm{R})+\\ &     \sum_{\beta \in B} \sum_{\bm{R}}C_{\mu \nu \beta}^{0,(\bm{R}'-\bm{R})} e^{-i\bm{k}\cdot(\bm{R}-\bm{R}')} \sum_{\bm{R}'} e^{i\bm{q}\cdot\bm{R}'}  f_{\beta}(\bm{r}-\bm{R}') \\ 
\end{aligned}
\end{equation}
and further, by defining $\bm{\Delta R} = \bm{R}-\bm{R}'$, we get
\begin{equation}
\begin{aligned}
  \chi^{*}_{\mu}(\bm{r},\bm{k})\chi_{\nu}(\bm{r},\bm{k}+\bm{q}) 
    & \approx \sum_{\alpha \in A} \sum_{\bm{\Delta R}} e^{i(\bm{k}+\bm{q})\cdot\bm{\Delta R}}  C_{\nu \mu \alpha}^{0,\bm{\Delta R}}\sum_{\bm{R}} e^{i\bm{q}\cdot\bm{R}}  f_{\alpha} (\bm{r}-\bm{R})  + \\ &   \sum_{\beta \in B} \sum_{\bm{\Delta R}} e^{-i\bm{k}\cdot\bm{\Delta R}} C_{\mu \nu \beta}^{0,-\bm{\Delta R}} \sum_{\bm{R}'} e^{i\bm{q}\cdot\bm{R}'}    f_{\beta} (\bm{r}-\bm{R}')  \;.  
\end{aligned}
\end{equation}
Defining
\begin{align}
   C_{\nu\mu\alpha}(\bm{k}) = \sum_{\bm{\Delta R}} C_{\nu\mu\alpha}^{0,\bm{\Delta R}}e^{i\bm{k}\cdot\bm{ \Delta R}} \;,
   \label{eq:kcoeff}
\end{align}
the periodic $\bm{k}$-dependent PADF equations read
\begin{equation}
\begin{aligned}
    \chi^{*}_{\mu}(\bm{r},\bm{k})\chi_{\nu}(\bm{r},\bm{k}+\bm{q}) \approx & \sum_{\alpha \in A} C_{\nu\mu\alpha}(\bm{k}+\bm{q}) f_{\alpha}(\bm{q},\bm{r})  +  \sum_{\beta \in B} C^{*}_{\mu\nu\beta}(\bm{k}) f_{\beta}(\bm{q},\bm{r}) \\ 
    & = \sum_{\gamma} C^{\mu\nu}_{\gamma}(\bm{k}+\bm{q},\bm{k})f_{\gamma}(\bm{q},\bm{r}) \;,
    \label{eq:dropPair0}
\end{aligned}
\end{equation}
where we have combined the separate summations over $\alpha$ and $\beta$ for notational convenience. In \cref{eq:dropPair0}, the structure of pair fitting is maintained: fit functions are still located on the same atoms as the basis set products. The notion of 'same atoms' implies that two atoms are considered equivalent as long as they are invariant by a direct lattice vector translation.
$P^{(0)}$ can be expressed in terms of Kohn--Sham (KS) states $\psi$ and KS eigenvalues $\epsilon$ as
\begin{align}
       P^{(0)}(\bm{r},& \bm{r}',i\omega)  =  \sum_{n,m}\sum_{\bm{k}+\bm{q},\bm{k}} \zeta_{\bm{k}}\zeta_{\bm{q}}  \label{eq:polrr} \\ \nonumber & 
       \frac{(o^{\bm{k}+\bm{q}}_{m}-o^{\bm{k}}_{n})\psi^{*}_{m}(\bm{r},\bm{k}+\bm{q})\psi_{n}(\bm{r},\bm{k})\psi^{*}_{n}(\bm{r}',\bm{k})\psi_{m}(\bm{r}',\bm{k}+\bm{q})}{\epsilon_{m}(\bm{k}+\bm{q}) - \epsilon_{n}(\bm{k}) - i\omega} \;,
\end{align}
where $m$ and $n$ denote band indices. We do not consider the case of partially filled bands and assume spin-compensation in this work. Therefore, the occupation factors $o$ are either $2$ for valence, and $0$ for conduction bands. The weights $\zeta_{\bm{k}}$ and $\zeta_{\bm{q}}$ account for the sampling of $K$ and $Q$ spaces over all Born--von Karman states in the first Brillouin zone and include the normalization factors of the Bloch summation functions.
Transforming \cref{eq:dropPair0} into the basis of KS states and adopting the usual convention of denoting virtual orbitals with the index $a$ and occupied ones with $i$,
\begin{equation}
    \psi^{*}_{a}(\bm{r},\bm{k})\psi_{i}(\bm{r},\bm{k}+\bm{q})  \approx \sum_{\gamma} \mathcal{C}^{ai}_{\gamma}(\bm{k}+\bm{q},\bm{k})f_{\gamma}(\bm{q},\bm{r}) \;,
    \label{eq:SolExp}
\end{equation}
and substituting eq~\eqref{eq:SolExp} in eq~\eqref{eq:polExpansion} we finally obtain the expression
\begin{align}
        P^{(0)}_{\gamma\gamma'}(\bm{q},i\omega)
    & = - 2 \sum_{\bm{k}}  \zeta_{\bm{k}} \zeta_{\bm{q}} 
 \sum_{i,a}\frac{\mathcal{C}^{ai}_{\gamma}(\bm{k}+\bm{q},\bm{k}) \mathcal{C}^{*ai}_{\gamma'}(\bm{k}+\bm{q},\bm{k}) }{ \epsilon_{a,\sigma}(\bm{k}) - \epsilon_{i,\sigma}(\bm{k}+\bm{q}) - i\omega} + c.c. \;.
\end{align}
Additionally, the Coulomb potential can be obtained as follows
\begin{equation}
\begin{aligned}
    \tilde{v}_{\gamma \gamma'}(\bm{q},\bm{q}')      & = \int \mathrm{d}r \mathrm{d}r' \frac{f^{*}_{\gamma}(\bm{q}',\bm{r}')f_{\gamma'}(\bm{q},\bm{r}) }{|\bm{r}-\bm{r}'|} \\
    & =  \int \mathrm{d}r \mathrm{d}r'\sum_{\bm{R},\bm{R}'} \frac{f_{\gamma}(\bm{r}'-\bm{R}')e^{-i\bm{q}'\cdot\bm{R}'}f_{\gamma'}(\bm{r}-\bm{R})e^{i\bm{q}\cdot\bm{R}} }{|\bm{r}-\bm{r}'|}  \\ 
    & =  \sum_{\bm{\Delta R}} \int \mathrm{d}r \mathrm{d}r' \frac{f_{\gamma}(\bm{r}'-\bm{R} -\bm{\Delta R})f_{\gamma'}(\bm{r}-\bm{R}) e^{-i\bm{q}'\cdot\bm{\Delta R}}  }{|\bm{r}-\bm{r}'|}  \sum_{\bm{R}} e^ {i(\bm{q}-\bm{q}')\cdot\bm{R}} \\
    & = N_{\bm{R}} \cdot \delta_{\bm{q},\bm{q}'} \cdot \sum_{\bm{\Delta R}} \big \langle f_{\gamma}(\bm{r}-\bm{\Delta R})  | \frac{1}{|\bm{r}-\bm{r'}|} | f_{\gamma'}(\bm{r}') \big \rangle e^{-i\bm{q}'\cdot\bm{\Delta R}}  \;,
    \label{eq:PeriodicFitR12Fit}
    \end{aligned}
\end{equation}
where the $N_{\bm{R}}$-factor can be set to one to yield the RPA correlation energy per unit cell.

\subsection{\label{sec:PPM}Periodic Projector Method}
When the PADF expansion is used, perturbation-theoretical methods relying on virtual orbitals often suffer from stronger numerical instabilities than methods utilising only occupied orbitals.\cite{Spadetto2023} To overcome this issue which arises from linear dependencies in the primary basis, we generalized the projector method of Ref.~\citenum{Spadetto2023} to periodic systems.
Starting from the $\bm{k}$-point specific overlap matrix $S_{\mu\nu}(\bm{k}) = \langle \chi_{\mu}(r,\bm{k}) | \chi_{\nu}(r,\bm{k})\rangle $
we build a projector $T(\bm{k})$ by diagonalizing $S$
\begin{align}
    S(\bm{k}) = U(\bm{k}) D(\bm{k}) U^{\dagger}(\bm{k})
\end{align}
and subsequently removing the subspace spanned by eigenvectors corresponding to eigenvalues smaller than a user-specified threshold $\epsilon_{d}$,
\begin{align}
\label{projector_threshold}
    R_{ij}(\bm{k}) = & \delta_{ij} \Theta(D_{ii}(\bm{k}) - \epsilon_{d}) \\ 
       \tilde{R}_{ij}  = &  \prod_{\bm{k}}  R_{ij}(\bm{k})\\ 
    T(\bm{k}) = & U(\bm{k}) \tilde{R}\, U^{\dagger}(\bm{k}) \,
\end{align}
where $\Theta$ is the Heaviside function. 
The $R(\bm{k})$ matrices are combined via matrix multiplication, hence, whenever a matrix element from any $R(\bm{k})$ is zero, it will be zero for the cumulative projector. This allows removing the same atomic specific linear combinations from every $\bm{k}$ point.
Eventually $T(\bm{k})$ is then used to regularize the KS orbital coefficients as
\begin{align}
    \tilde{c}^{i}_{\mu}(\bm{k}) =  \sum_{\nu} T_{\mu\nu}(\bm{k}) c^{i}_{\nu}(\bm{k}) \;.
    \label{eq:pmeig}
\end{align}
These coefficients are then used to transform \cref{eq:dropPair0} to \cref{eq:SolExp}. Numerical issues can become more pronounced with increasing system and basis set size, necessitating the use of this projector method. Of course, this issue is strongly related to the quality of the auxiliary fit set. When large fit sets are used, a small $\epsilon_{d}$ will suffice, while a smaller fit set necessitates a larger value of $\epsilon_{d}$\cite{Spadetto2023}.
\usetikzlibrary{shapes.geometric, arrows, calc}

\definecolor{qwcolor}{RGB}{255,245,151}
\definecolor{qqcolor}{RGB}{151,203,255}

\tikzset{
  basic/.style={
    draw,
    rectangle,
    text width=2cm, 
    font=\sffamily,
    fill=green!30
  },
  style root/.style={
    basic, 
    rectangle,
    thin, 
    align=center,
    fill=green!30
  },
  style omega/.style={
    basic, 
    rectangle, 
    thin,
    align=center, 
    fill=red!60,
    text width=10mm,
    anchor=north
  },
  style omega/.style={
    basic, 
    rectangle, 
    thin,
    align=center, 
    fill=red!40,
    text width=8mm,
    anchor=north
  },
   style qq/.style={
    basic, 
    rectangle, 
    thin,
    align=center, 
    fill=qqcolor,
    text width=10mm,
    anchor=north
  },
  style qwloop/.style={
    basic, 
    rectangle, 
    thin,
    align=center, 
    fill=qwcolor,
    text width=8mm,
    anchor=north
  },
  child node/.style={
    basic, 
    circle, 
    thin,
    align=center, 
    fill=green!60,
    text width=1mm,
    anchor=north
  },
  every child node/.style={child node}
}
\section{\label{sec::implementation}Implementation}
\subsection{\label{sec:damping}Restricted Bloch summation}

The periodic RI-RPA equations \cref{eq:dropPair0,eq:blochsumdefs,eq:kcoeff} but also the $\bm{q}$-dependent Coulomb overlap integrals \cref{eq:PeriodicFitR12Fit}, formally rely on Bloch summations to obtain a mapping between two infinite spaces. 
In practice, the reciprocal space grid will however be a finite sampling of the infinite number of $\bm{k}$-points in the first Brillouin zone prescribed by the Born--von Karman boundary conditions. This unavoidable limitation demands additional considerations on the representation of such quantities. 
Since the set of Bloch states is obtained from its real-space analog via a discrete Fourier transform of the unit cell coordinate $\bm{R}$, a finite $\bm{k}$-grid implies that only a limited amount of unit cells can be considered. In other words, for a regular $\bm{k}$-sampling along the reciprocal lattice vectors, there will be a maximum representable distance in real space. Starting from the minimum increment between $\bm{k}$-points $ \text{min}[ k^{x}_l - k^{x}_n] = q^{x}_{min}$ along each reciprocal lattice vector direction $i$, the 'Nyquist' distance ${R^{i}_{max}}$, on the corresponding real lattice vector direction, is defined by
\begin{align}
q^{i}_{min} \cdot R^{i}_{max} = 2 \pi \;.
\label{eq:maxgrid}
\end{align}
% We refer to Appendix~\ref{sec:maxlen} for further discussions of this relation.
${R^{i}_{max}}$, defines an $n$-dimensional parallelepiped ($n=3$ for bulk). Within this parallelepiped, a new real space grid is defined, containing as many unit cells as the number of $\bm{k}$-points with coordinates $R_{x}$. To ensure that the Fourier transform is invertible, it is necessary to restrict every Bloch summation to cells lying only within the grid induced by the Nyquist vectors ${R^{i}_{max}}$. Beyond ${R^{i}_{max}}$, every function represented through $Q_{G}$ is not properly defined, unless it reaches zero within the limits of ${R^{i}_{max}}$. 
If not, trying to represent them with a unsuitable $Q_{G}$ would introduce spurious long-range effects. On the other hand, due to the non-invertibility of the Fourier transform, functions that do not decay to zero within $R^{i}_{max}$ would artificially repeat if transformed back to real space beyond ${R^{i}_{max}}$. This induces undesirable boundary effects for non-converged $\bm{k}$-grids. To ensure a proper decay within ${R^{i}_{max}}$, Spencer and Alavi suggested to attenuate the Coulomb potential by setting $v(\bm{r}_{1},\bm{r}_{2})$ to zero beyond ${R^{i}_{max}}$.\cite{Spencer2008} Here, we instead introduce a spherical Fermi-Dirac function,
\begin{equation}
    \theta(|\bm{r}_{2}-\bm{r}_{1}|, r_{0}, \beta) = \frac{1}{1+e^{\beta(|\bm{r}_{2}-\bm{r}_{1}|+r_{0})}}
\end{equation}
and damp the Coulomb potential according to
\begin{align}
   & v_{\theta}(\bm{r}_{1},\bm{r}_{2}) = v(\bm{r}_{1},\bm{r}_{2})  \theta(|\bm{r}_{2}-\bm{r}_{1}|,r_{0}, \beta) \;. 
\end{align}
Similar schemes to damp the Coulomb potential have previously been used in periodic HF calculations.\cite{Sundararaman2013}

\begin{figure}
    \centering
    \includegraphics[height=0.35\textwidth,angle=270,origin=c]{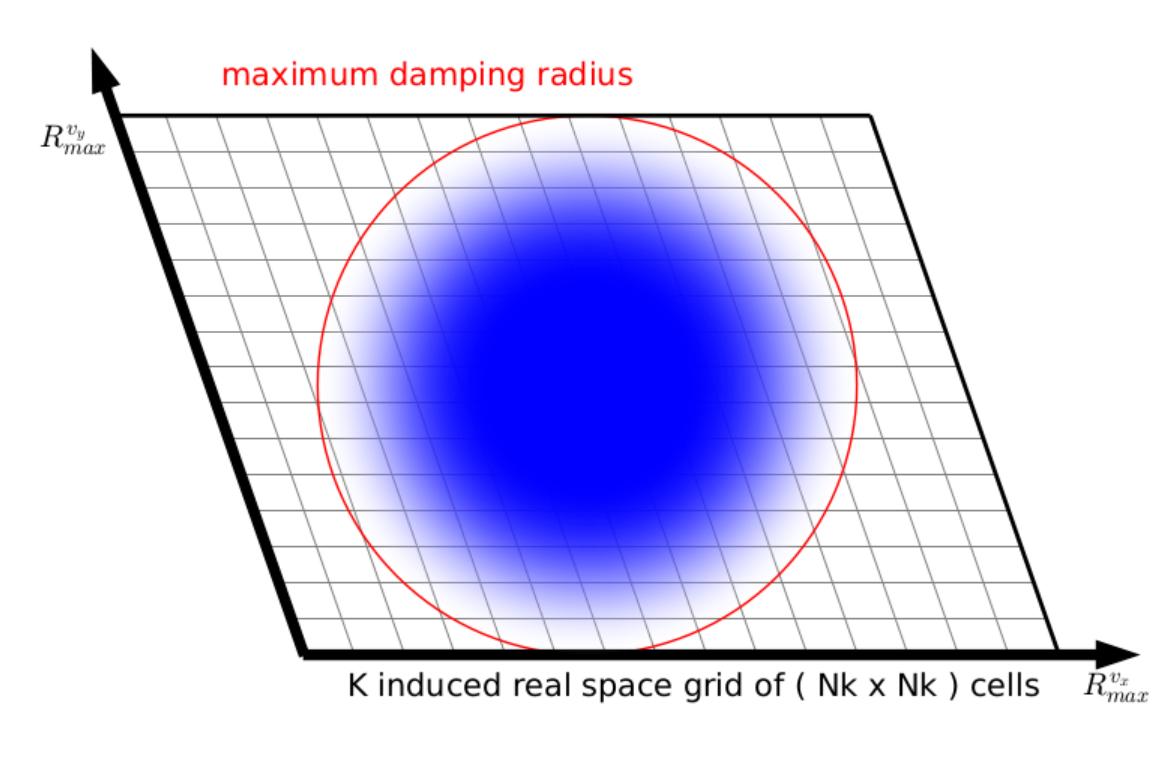}
    \vspace{-40pt}
    \caption{Pictorial representation of the $R^{x}$ grid. Each grey cell represents a primitive cell. The whole grid contains $15\times15$ unit cells, which can be represented by $15\times15$ points in the k grid. The red circle represents the maximum circle that can fit in this grid and the blue shadow is the damping function for the Coulomb potential.}
    \label{fig:nyqst}
\end{figure}

As illustrated in Fig.~\ref{fig:nyqst}, the parameters of $\theta$ directly depend on the chosen $\bm{k}$-grid. If not specified otherwise, we choose the damping radius $r_{0}$ as $50 \%$ of $R_{c}$, the radius of the largest circle which fits the parallelepiped defined by $R_{max}^{i}$. The decay parameter $\beta$ is chosen to reduce the damping function to $0.1\%$ within a distance of 1.4 $ r_{0}$. In general, we will refer to the practice of damping the Coulomb potential depending on the $\bm{k}$-grid as AUTO damping. In this way, increasing the $\bm{k}$-grid will increase $r_{0}$ and $\beta$. In the limit of an infinite $\bm{k}$-grid, we would sample the infinite number of unit cells.

\subsection{\label{sec32}Dual grids for enhanced sampling around the $\Gamma$-point}
\label{sec:enhqgrids}

\begin{figure}[hbt!]
    \centering
    \begin{minipage}{0.45\textwidth}
      \begin{tikzpicture}
         \node at (0.42\textwidth,0.43\textwidth) { \includegraphics[width=\textwidth]{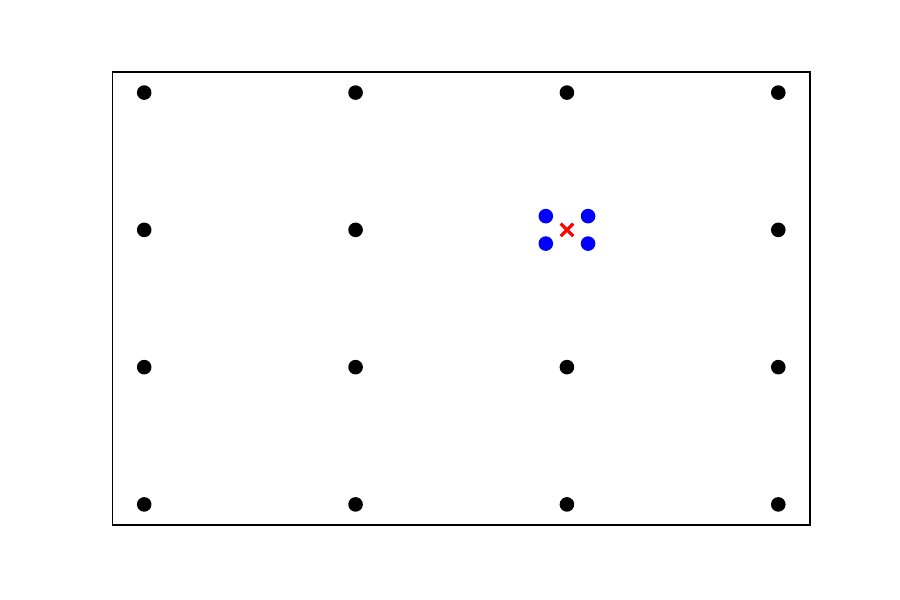}};
         \node at (0.02\textwidth,0.74\textwidth) {\textbf{A}};
  \end{tikzpicture}
    \end{minipage}
    \begin{minipage}{0.45\textwidth}    
          \begin{tikzpicture}
         \node at (0.42\textwidth,0.43\textwidth) {     \includegraphics[width=\textwidth]{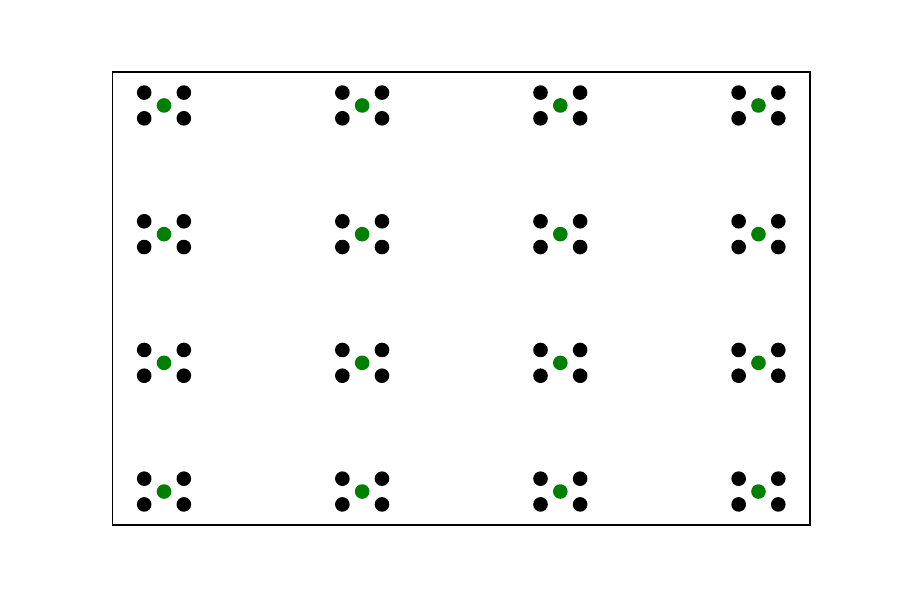}};
         \node at (0.02\textwidth,0.74\textwidth) {\textbf{B}};
  \end{tikzpicture}
    \end{minipage}
    
    \caption{In (A) we show the $Q_{G}$ grid. The additional sampling near $\Gamma$ is represented in blue, the neglected $\Gamma$ point with a red cross, and the other regular points in black. In (B) we show the $K_{G}$ grid. The regular set $K_{reg}$, which is sampled at first is represented in green, in black all the other points which are added accordingly to $Q_{G}$. Using ~\eqref{eq:travelK} from green points in (B) with black points in (A) will transform to a green point $\bm{k}_{j}$. The additional blue points in $Q_{G}$ create the surrounding black points in (B).
   }
    \label{fig:q_Kgrids}
\end{figure}

Due to the slow convergence of RPA correlation energies, regular sampling of the first Brillouin zone often requires unpractically large $\bm{k}$-grids to reach satisfactory accuracy.\cite{Huser2013a, Pela2024, Guandalini2023a} 
This is caused by the divergence of the Coulomb potential at the $\Gamma$-point and can be mitigated by truncating the Coulomb interaction as described in the previous section. For a non-converged $\bm{k}$-grid, such a truncation would however be artificial and undesirable.
To overcome this issue, we enhance the sampling of $Q_{G}$ around $q\equiv \Gamma$ by generating a regular grid in $\bm{q}$, removing $\bm{q}\equiv \Gamma$, and adding additional points in the vicinity of $\Gamma$ as can be seen  Fig.~\ref{fig:q_Kgrids}A. This approach is inspired by the staggered mesh method of Lin and coworkers.\cite{Xing2021, Xing2022}
Differences lie in the fact that while we are using the same grids for occupied and virtual orbitals, we have an independent $\bm{q}$-grid augmenting our regular $\bm{k}$-grid. This allows for introducing a flexible displacement parameter in $\bm{q}$ to treat the integrable divergence at the $\Gamma$-point.

Considering the reciprocal lattice vectors $\bm{b}_{i}$ as basis for the reciprocal space, these additional points have coordinates $\bm{x} = \sum_{i} \pm d \bm{b}_{i}$, where $\bm{x}$ is position of the additional point in $Q_{G}$. According to the dimension of the system, we add 2,4, and 8 points for respectively 1D, 2D and 3D systems. 
With this procedure, we sample $\bm{k}$-space, by generating a regular k-grid $K_{reg}$ first and then combining it with the increments $Q_{G}$ as
\begin{align}
    \bm{k}_{j} + \bm{q} = \bm{k}_{i} \;.
    \label{eq:travelK}
\end{align}
In this way, all $\bm{k}_{i}$ which augment $K_{reg}$, will be used to generate the final $K_{G}$, which is schematically presented in Fig.~\ref{fig:q_Kgrids}B. 

With the right choice of $Q_{G}$, the size of $K_{G}$ increases linearly with  $Q_{G}$, because many target points $\bm{k}_{i}$ will already be part of the regular set $K_{reg}$. The additional sampling increases the size of the $K_{reg}$ grid exactly by a multiplicative factor of $N_{add}+1$.
For instance, for a 2D system, where 4 additional points are added in $Q_{G}$ around $\Gamma$, a regular grid $K_{reg}$ of $10\times 10$ $\bm{k}$-points, results in a $K_{G}$ of $(10\times10) \times 5$. 
This scaling however only affects the memory demands of the calculations. As can be seen from \cref{eq:polrr}, the computational effort to evaluate the polarizability in reciprocal space scales as the product of the $\bm{k}$ and $\bm{q}$ integration grids, which is $K_{reg}\times Q_{G}$.

In the worst-case scenario, using other choices of $Q_{G}$, such as generic non-regular grids, can make the size of $K_{G}$ scale quadratically with respect to $Q_{G}$. From \cref{eq:travelK}, it is evident that this occurs when none of the $\bm{k}_{j}$ points belong to the previous $K_{reg}$ sampling.
Other samplings around $\Gamma$, improving on a regular grid, also make the size of $K_{G}$ scale linearly with respect to $Q_{G}$, but the number of additional points influences the prefactor of the memory scaling.
Since now the minimum increment between two $\bm{k}$-points is determined by the distance of the additional sampling around $q = \Gamma$, also the radius of the Coulomb potential damping increases.
In particular, for the results that will be presented further in this article, the distance $d^{x} = \frac{1}{10} d^{x}_{reg}$ where $d^{x}_{reg}$ is the directional increment between the initial regular grid in $\bm{q}$. 

\subsection{RPA Algorithm}

\begin{algorithm} 
\caption{Algorithm for the evaluation of the RPA correlation energy}
\label{alg:loop}
\begin{algorithmic}[1]
\Require{ $\mathcal{C}^{ia}_{\alpha}$ , $\epsilon_{n}$ } 

\For{$q \gets q_{1}$ to $q_M$ (subdivide in $N_{g.q}$ groups) }        
\State eval. $V^{0.5}(q)$
\For{$\omega \gets \omega_{1}$ to $\omega_N$ (subdivide in $N_{g.\omega}$ groups)}   
    \For{ $k_{1}$ and $k_{2}$ if $k_{1}-k_{2} \equiv q$}
     \State {eval. $\mathcal{C}^{*k_{1}k_{2}}_{a,i,\beta}$}
     \State {eval. $P_{k_{1},k_{2}}(i\omega)$}
     \State {eval. $\mathcal{C}^{*k_{2}k_{1}}_{a,i,\beta}$}
     \State {eval. $P^{\dagger}_{k_{2},k_{1}}(\omega)$}
     \State {$P_{q}(i\omega) = P_{q}(i\omega) + \zeta_{k} \big [ P_{k_{1},k_{2}}(\omega)+P^{\dagger}_{k_{2},k_{1}}(i\omega) \big ] $}
    \EndFor
     \State {$Z_{\omega,q} = V^{0.5}_{q} P_{q}(i\omega) V^{0.5}_{q}$}
     \State {$\varepsilon_{\omega,q} = \big[\vspace{1pt} \log (\det(\mathds{1}-Z_{\omega,q})) - \Tr(Z_{\omega,q}) \vspace{1pt} \big ]$}
     \State $e_{\omega} = e_{\omega} + 2 \zeta_{q} \varepsilon_{\omega,q} $        
\EndFor     
\EndFor

    \State $E_{_{RPA}} = \sum_{\omega} e_{\omega} \zeta_{\omega}$
\end{algorithmic}
\end{algorithm}

\newcommand{\lqw}{ $P_{q}(\omega)$, $Z_{\omega,q}$, $\varepsilon_{\omega,q}$}
\newcommand{\sw}{style omega}
\newcommand{\ewq}{$e_{\omega}$}
\newcommand{\nqw}{$n_{q,\omega}$}
%\resizebox{\textwidth}{!}{%

%\itc{\onecolumn} 
%\begin{center}
    
\begin{figure}[hbt!]
\begin{center}
\scriptsize
\begin{tikzpicture}
[level distance=10mm,
   level 1/.style={sibling distance=12mm},
   level 2/.style={sibling distance=5mm},
   level 3/.style={sibling distance=5mm},
   level 4/.style={sibling distance=5mm},inner frame sep=20pt,framed,background rectangle/.style={thick,draw=black}]
\coordinate node [style root] (00){$N_{p}$} 
        child { node [\sw]  (4) {\nqw}  
                child {  node [style qwloop](42){\lqw} 
                    child {node [\sw](AA) {\ewq} } }  }
            child {node [\sw]  (5) {\nqw}  
                child {  node [style qwloop](52){\lqw} 
                    child {node [\sw](AB) {\ewq} 
                        child {node [style qq] (A) {combine ${\omega,q}$} }}}}
            child {node [\sw]  (6){\nqw}   
                child {  node [style qwloop](62){\lqw} 
                    child {node [\sw](AC) {\ewq} } }};

\node (RR) [label,text width=6cm, left = 2cm of  00  ]  {Sequential context } ; 

\node (BB) [label, text width=6cm, below = 0.6cm of RR]  { Split different $N_{p}$ contexts in $N_{g.\omega} \cdot N_{g.q} $ groups };
\node (cc)  [label, text width=6cm, below  =  0.8cm of BB ] {Evaluate ($\bm{q}$,$\omega$) specific intermediates};
\node (DD)[label, text width=6cm,  below  =  0.8cm of cc  ] {Integrate partially over $\bm{q}$ };
\node[label, text width=6cm,  below  =  0.8cm of DD  ] {Combine partial integrals in $\bm{q}$ and integrate on $\omega$};

\draw [-] (42) -- (AA);
\draw [-] (52) -- (AB);
\draw [-] (62) -- (AC);

\draw [-] (AA) -- (A);
\draw [-] (AB) -- (A);
\draw [-] (AC) -- (A);

\end{tikzpicture}
\end{center}
    \caption{Schematic overview of the parallelization strategy adopted in our implementation.}
    \label{fig:parallelization}
\end{figure}

Here, we discuss the design of our RPA algorithm which is summarized in algorithm~\ref{alg:loop}. The decisive part is the calculation of the RPA integrand $\varepsilon_{\omega,q}$ for all $\bm{q}$ and $\omega$ in their respective grids which enters eq.~\eqref{working-equation} through
\begin{align}
\label{mainLoopEquation}
    E^{RPA}_{corr} \approx  \sum_{\omega} \sum_{\bm{q} \in Q_{G}}  \varepsilon_{\omega,\bm{q}} \zeta_{\bm{q}} \zeta_{\omega} \;,
\end{align}
where $\zeta_{\bm{q}}$ and $\zeta_{\omega}$ denote the integration weights for $\bm{q}$ and $\omega$-integration, respectively.
In our implementation, the $\bm{q}$-integration for a specific $\omega$ is achieved through subsequent summations at every cycle, obtaining in the end the intermediate $e_{\omega}$. The integration of $e_{\omega}$ is performed at the end of both loops.

To parallelise \eqref{mainLoopEquation} efficiently, the loops over $\omega$ and $\bm{q}$ group all the concurring processes $N_p$ in smaller ScaLapack contexts. This technique allows each ScaLapack context to be composed of $n_{\bm{q},\omega} = \nicefrac{N_p}{N_{g. \omega}N_{g. \bm{q}}}$ processes where each handles a portion of the whole workload, consisting of couples of $\bm{q}$ and $\omega$.
Each group will have a separate ScaLapack context to enable distributed algebra.
In practice, considering sequential routines proportionally more efficient than their distributed version, one would prefer to have only one single process per group, without actually adopting any parallelisation in the algebra. However, this is not always achievable due to memory constraints, since storing all the intermediate matrices at the same time can become prohibitive. Apart from the communication that occurs within ScaLapack, the only communication that is needed consists in gathering $e_{\omega}$, which later is integrated only in the main node with the respective weights. The procedure is illustrated in Fig.~\ref{fig:parallelization}.

The fit coefficients are transformed to the MO basis on the fly when needed for the corresponding polarizability terms. Therefore, some of these transformations have to be performed multiple times. However, the alternative of storing them all, even on disk, is prohibitive due to memory constraints. Nevertheless, for some architectures with faster read and write to disk, or for small calculations in which the memory is not a constraint, these become viable options. For larger systems, storing the PADF fit coefficients becomes the memory bottleneck.
\section{\label{sec::results}Results}
In the following, we report tests of several aspects of our RPA implementation. In all calculations, we employ STO-type basis sets represented in the NAO-form. These range from double-$\zeta$ (DZ) to quadruple-$\zeta$ quality (QZ), referred to as DZP, TZ2P, and QZ4P, respectively,\cite{vanLenthe2003} where the suffix $x$P indicates the number of polarization functions. We note that some integrals, such as kinetic energy integrals or overlap integrals, could in principle be evaluated analytically for STOs. However, the computational overhead of evaluating them numerically when they are represented as NAOs is completely negligible. All calculations used a modified Gauss--Legendre frequency grid of 32 points, according to the prescription of Ref.~\citenum{Fauser2021}. All other computational details are provided in connection with the rest of the results. 

\subsection {Coulomb potential truncation}

\begin{figure}[hbt!]
    \centering
    \includegraphics[width=0.7\textwidth]{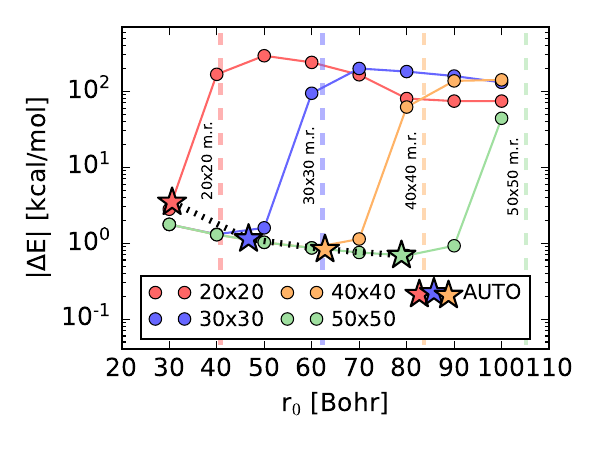}
    \caption{Effect of Fermi--Dirac damping on RPA@PBE correlation energy convergence for hexagonal boron nitride (STO-DZ basis) The y-axis shows the absolute deviation from a reference RPA energy, extrapolated via an inverse-linear fit in the infinite $Q_G$ limit using the \texttt{AUTO} damping scheme. The x-axis varies the Fermi--Dirac damping parameter $r_0$. Colors denote different $Q_G$ grids, with vertical lines indicating the largest inscribed circle radius in the corresponding $R_x$ grids. All curves use the same $\beta$ parameter, as defined along with the \texttt{AUTO} damping settings in Sec.~\ref{sec:damping}.}
    \label{fig:hbn_auto}
\end{figure}

As a first test of our implementation, we assess how the damping of the Coulomb potential affects the accuracy of our results, and how it influences convergence of the RPA correlation energies with respect to the size of the $\bm{k}$-grid for two-dimensional hexagonal boron nitride (h-BN). 
Fig.~\ref{fig:hbn_auto} shows the absolute errors of the RPA correlation energy for different $r_{0}$ in the Fermi-Dirac damping function for different regular $\bm{k}$-grids, distinguished by different colours. We have chosen the infinite $\bm{k}$-grid limit, obtained from fitting the AUTO damping scheme with an inverse linear fit with respect to the number of $\bm{k}$-points as reference value. We have used a constant decay parameter $\beta$, obtained according to the rules depicted in Sec.~\ref{sec:damping} in all calculations.
The error in the RPA correlation energy decreases with the number of $\bm{k}$-points, as long as the damping radius $r_0$ is smaller than the radius of the circle confined by the parallelogram defined by $R^{max}$. Larger values of $r_0$ lead to uncontrolled errors in the correlation energy. 

Moreover, as long as $r_0$ is smaller than  $R^{max}$, we observe convergence of the RPA correlation energy to a short-range separated value. For instance, choosing a fixed damping radius of $r_0 = 40$~Bohr, we obtain almost the same RPA correlation energy with the $30\times30$ and $50\times50$ $\bm{k}$-grids. A fixed damping radius would correspond to a fixed short-range separated RPA calculation. This is not desirable, since RPA correlation is important in the long range.\cite{Bruneval2012b} 
 Moreover, the artificial convergence comes from restricting the RPA correlation energy to the short range, leading to potentially large errors in absolute correlation energies.

The automatic damping we use in this work avoids such an artificial truncation of the Coulomb potential outside the finite resolution prescribed by the finite $\bm{k}$-grid. The stars in Fig.~\ref{fig:hbn_auto} show the convergence of the RPA correlation energy with respect to the $\bm{k}$-grid and associated damping distance to the true undamped result. For this test, we used a damping distance $r_{0}$ equal to 75\% of $R_{c}$, with a decay speed that reached a damping factor of 0.1\% at a distance of 1.3$r_{0}$.
Lastly, we bring to the attention the large deviations from the fully converged result, which in the best cases is of the order of 1 kcal mol$^{-1}$. Leaving out that we are looking at an absolute energy and that energy differences will be subjected to error cancellation, the root cause is the difficulty of regular grids to treat the integration at the $\Gamma$ point.
In section~\ref{sec43}, we will show how the use of dual grids, described in section ~\ref{sec32} improves the $\bm{k}$-point convergence.

\subsection{Damping distance effects on fluoro-polyacetylene }

\begin{figure}[hbt!]
    \centering
    \begin{minipage}[t]{0.45\textwidth}
    \begin{tikzpicture}
         \node at (0.42\textwidth,0.43\textwidth) {    \includegraphics[width=\textwidth]{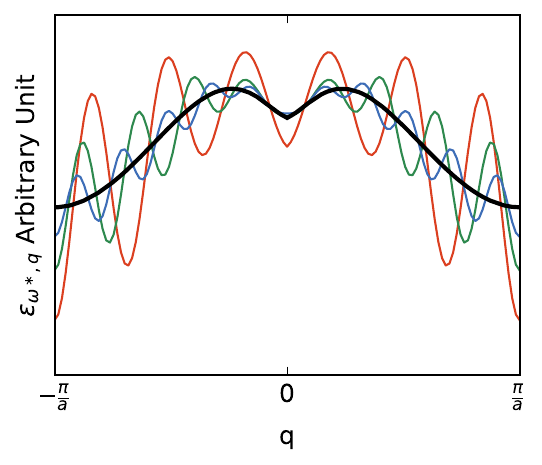}};
         \node at (0.08\textwidth,0.785\textwidth) {\textbf{A}};
     \node at (0.42\textwidth,0.3\textwidth) { \includegraphics[width=\textwidth]{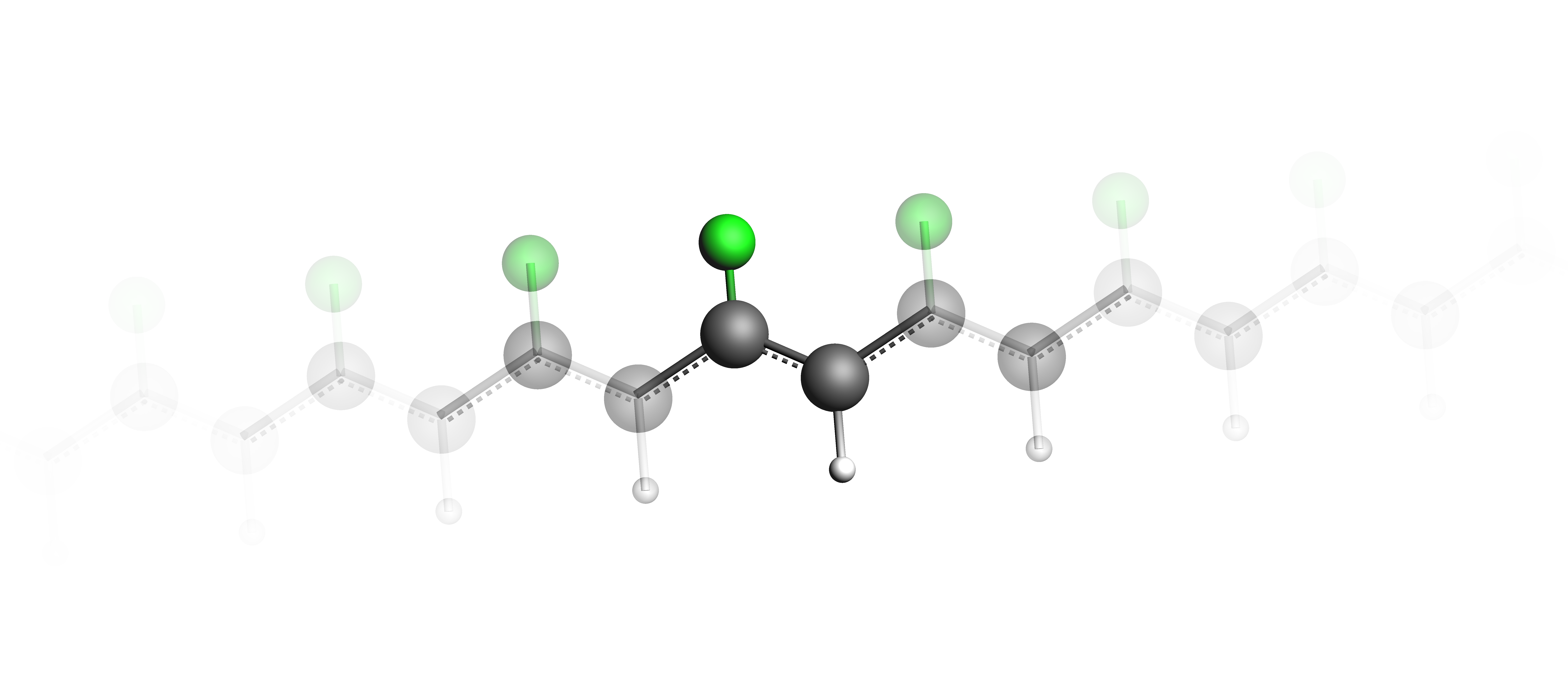}};
  \end{tikzpicture}
    \end{minipage}
    \begin{minipage}[t]{0.45\textwidth}
    \begin{tikzpicture}
        \node at (0.42\textwidth,0.43\textwidth) {    \includegraphics[width=\textwidth]{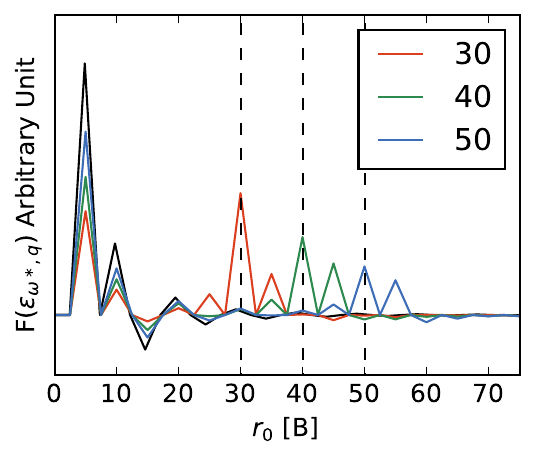}};
        \node at (0.08\textwidth,0.785\textwidth) {\textbf{B}};
  \end{tikzpicture}
    \end{minipage}
      \caption{in (A) Fluoro-polyacetylene geometry with its RPA correlation energy integrand $\epsilon_{\omega,q}$ for a specific imaginary frequency, $i\omega = 8$ Hartree, using the DZ basis set of Band. The different numbers in the legend denote the different values of $r_0$ in Bohr. In (B) Its Fourier analysis for the same frequency and same basis. The black line represents a calculation performed with 1000 $\bm{k}$-points and AUTO range separation}
    \label{fig:rsepswq}
\end{figure}
%927529
To further study the effects of the damping of the Coulomb potential on RPA correlation energies we analyzed the RPA integrand  $\varepsilon_{\omega,q}$ of fluoro-polyacetylene (whose structure is shown in Fig.(~\ref{fig:rsepswq} ) for different Fermi-Dirac damping distances $r_{0}$ using a decay parameter $\beta$ reaching a damping of 0.1\% at distance of 1.3 $r_{0}$ using regular 1D grids $K_{G}$ and $Q_{G}$ of $128$ $\bm{k}$-points. 
The results, displayed in Fig. \ref{fig:rsepswq}, demonstrate that the damping of the Coulomb potential, or, more generally, the effect of a finite grid of unit cells in real space, do not solely affect the integrand at the $\Gamma$-point ($\bm{q} \equiv 0$) but modifies $\varepsilon_{\omega,q}$ globally. 

We notice that $\varepsilon_{\omega,q}$ tends to converge to a definite value at the $\Gamma$-point for increasing damping distances, while the behaviour away from $q \equiv \Gamma$ point is more subtle. Guided by the oscillating trend of the curves in the left panel of Fig.~\ref{fig:rsepswq}, we performed a Fourier analysis of the same function presented in the right panel of Fig.~\ref{fig:rsepswq}, where, given the parity of the function, only the positive axis is shown. The different lines exhibit unique spikes in increasing order, according to the respective damping radii.

\begin{table}[hbt!]
    \centering
    \begin{tabular}{c|ccc}
        $r_{0} $ & $E_{RPA}( 2.47 \si{\angstrom} ) $   & $E_{RPA}( 2.67 \si{\angstrom} ) $   & $\Delta  $   \\
        \hline

10	&	-333.408	&	-339.461	&	-6.052	\\
20	&	-331.902	&	-338.277	&	-6.374	\\
30	&	-331.731	&	-338.145	&	-6.414	\\
40	&	-331.694	&	-338.113	&	-6.418	\\
50	&	-331.683	&	-338.100	&	-6.417	\\

    \end{tabular}
    \caption{RPA correlation energies for fluoro-polyacetylene (kcal  mol$^{-1}$) using different bond lengths between the two carbon atoms (indicated in the parentheses), and different damping distances $r_0$ (first column, in Bohr) The RPA contribution to the energy difference between the equilibrium and stretched geometries (kcal  mol$^{-1}$) is reported in the last column, using the DZ basis set.}
    \label{tab:c2hflcdeltarpa}
\end{table}

Even though the functions $\varepsilon_{\omega,q}$ are very different for different damping distances, the RPA results reported in Table~\ref{tab:c2hflcdeltarpa} remain stable.
This suggests that for this system, to some extent, the effect of the damping cancels out in the oscillations shown in Fig.~\ref{fig:rsepswq}.
It is clear that the damping, which undoubtedly guarantees a regular convergence to the fully periodic RPA Correlation energy, affects the RPA integrand. As shown by the Fourier analysis, this is mitigated only when the damping parameter goes to infinity, which then comes with the known problem of the $\Gamma$-point divergence of the Coulomb potential.\cite{Ren2021} Damping of the Coulomb potential facilitates convergence to the limit of infinite $\bm{k}$-grid sampling, but it does not necessarily guarantee fast convergence. As shown in the next section, the combination with the dual grid approach leads to fast $\bm{k}$-grid convergence. 

\subsection{\label{sec43}Validation of damping approach and the reciprocal space convergence}

\begin{figure}[hbt!]
\centering
\includegraphics[width=\linewidth]{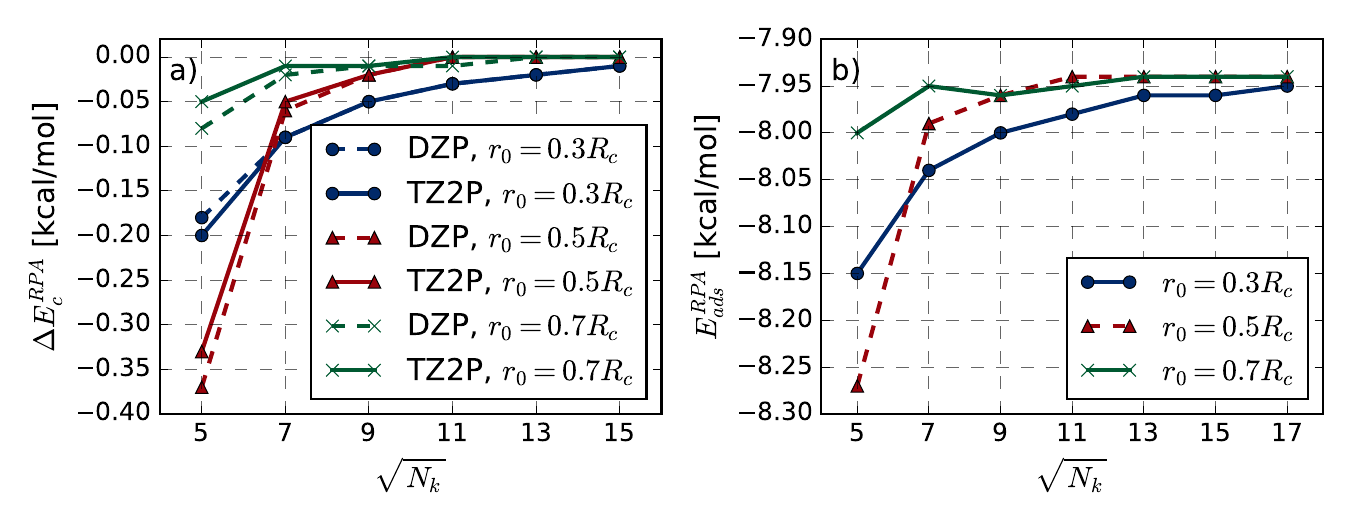}

\caption{Absolute deviation of the RPA contribution to the adsorption energy with respect to ts values calculated with a $17 \times 17$ $\bm{k}$-mesh, for different decay parameters and basis sets. b) RPA contributions to the interaction energy for different decay parameters. In all calculations, we set a damping factor of 0.1\% at $1.4r_{0}$}
\label{fig:devMonoMgOCO}
\end{figure} 

\begin{figure}
    \centering
\includegraphics[width=0.5\linewidth]{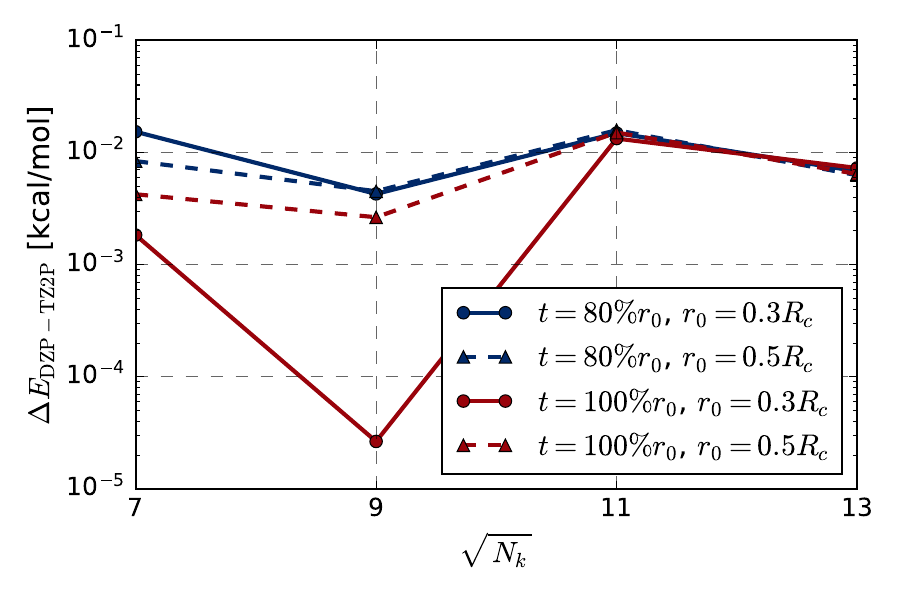}
\caption{Differences between the RPA contributions to the adsorption energy for the h-BN bilayer for different $\bm{k}$-point meshes relative to the values obtained with a $15 \times 15$ grid calculated with the DZP and TZ2P basis sets for different Coulomb damping parameters. The $t$ parameter prescribes a damping factor of 0.1\% at $r_{0} + 0.5t$.}
\label{fig:convDecayBN}

\end{figure}
To validate the combined use of damping and dual-grid techniques, we calculate the MgO--CO adsorption energy for a single MgO layer at 100\,\% coverage with a Mg--C distance of 2.479\,\r{A}, as well as the interaction energy of an AA$'$-stacked h-BN bilayer at an interlayer distance of 3.45\,\r{A}~\cite{Leconte2017}. Enhanced grids were used, adding sampling at $\nicefrac{1}{10}$ of the regular $\bm{k}$-grid step.

Fig.~\ref{fig:devMonoMgOCO}a) shows the CO adsorption energy on the MgO monolayer for different $\bm{k}$-meshes relative to its value calculated with a $17 \times 17$ $\bm{k}$-mesh. The data demonstrates that $\bm{k}$-point convergence is both rapid and that the rate of convergence is independent of the basis set: lines of the same color (DZP vs.\ TZ2P) for the same decay parameter $r_0$ nearly overlap. This observation is consistent with previous plane-wave results\cite{Jauho2015}. Fig.~\ref{fig:devMonoMgOCO}b) shows again the RPA contribution to the CO adsorption energy on the MgO monolayer for all different $\bm{k}$-meshes. The data clearly shows that the same thermodynamic limit is reached, independent of the precise value of $r_0$. The convergence rate is excellent, and independent of $r_0$, already a $9 \times 9$ $\bm{k}$-mesh is sufficient to converge the RPA contribution to the adsorption energy with 0.05 kcal/mol. Convergence becomes faster when $r_0$ is chosen as a larger fraction of $R_c$, since more and more of the Coulomb potential is accounted for for a given $\bm{k}$-grid ($r_0 = R_c$ would correspond to zero damping). Tables~S4~and~S6 in the supporting information demonstrate that absolute correlation energies converge equally fast.

For the h-BN bilayer, Fig.~\ref{fig:convDecayBN} shows the differences in RPA adsorption energies (DZP vs.\ TZ2P) at various $\bm{k}$-grids relative to a $15 \times 15$ reference. All values are very close to zero, demonstrating that convergence is independent of the basis set. In all of the following calculations, we use a damping of $r_0 = 0.5R_c$, since it seems to guarantee both a rapid and smooth convergence to the thermodynamic limit.

\subsection{Parallel performance}

\begin{figure}[H]
    \centering
    \includegraphics[width=0.45\textwidth]{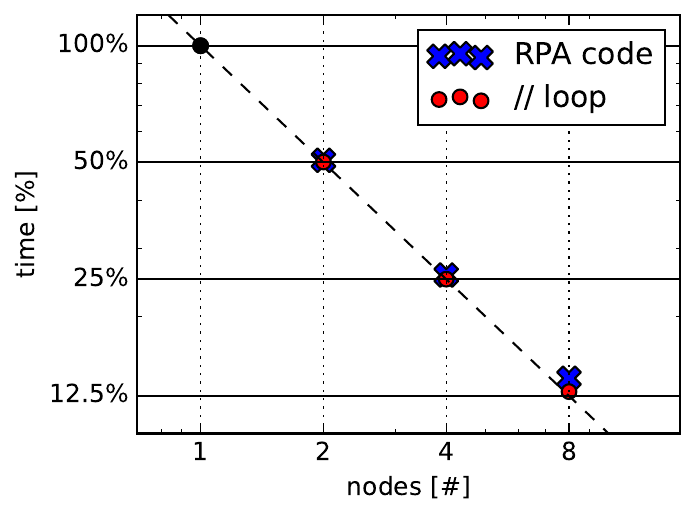}
    \caption{Percentage speed up of RPA correlation energy calculation for fluoro-poliacetylene.}
    \label{fig:c2hftime}
\end{figure}

In this section, we discuss the parallel efficiency of our algorithm. All calculations discussed here have been performed on AMD Genoa 9654 nodes with 192 cores, 2GB of DRAM per core, and 2.4GHz clock speed.

In \cref{fig:c2hftime}, we report the strong scaling of the algorithm through a test on a 1D fluoro-polyacetylene chain. It can be seen that the algorithm scales almost perfectly with the number of nodes. The timings for the computationally most intensive part of the algorithm, involving loops over $\bm{q}$ and $\omega$, are in almost perfect agreement with the theoretical speedup, represented through the dashed line.
Nevertheless, we find a small deviation of increasing relative magnitude. This is likely due to a non-perfect balance with respect to the $\bm{q}$ variable, which involves evaluating the Coulomb potential. In practice, if a specific $\bm{q}$ point is duplicated over different nodes, multiple evaluations of the potential for the same point are performed.
For the overall speedup of the algorithm, the considerations made for the $\bm{q}$, $\omega$ loop apply as well, but the deviations discussed before are larger. This is due to the necessary preparation steps before the start of the main RPA loop. These steps are carried out separately on each node to avoid the transfer of very large tensors.

\begin{figure}[hbt!]
    \centering
    \includegraphics[width=0.45\textwidth]{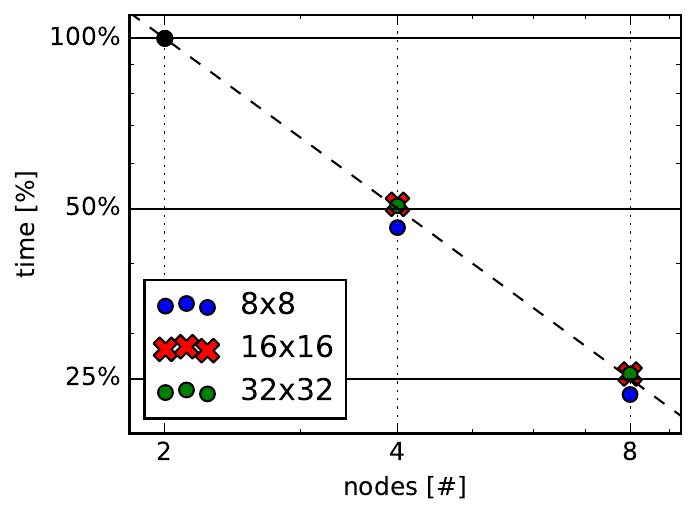}
    \caption{Percentage speed up of RPA correlation energy calculation by reducing the $K_{G}$ size h-BN.}
    \label{fig:hBN}
\end{figure}

\begin{table}[hbt!]
    \centering
    \begin{tabular}{c
    S[table-format=3.0]
    S[table-format=4.0]
    S[table-format=4.0]
    S[table-format=4.0]}
    \toprule
          \# nodes & {$\sqrt{K_{G}}$} & {$\bm{q}, \omega$-loop [s]} & {RPA} [s] & {RPA + SCF  [s]}\\
   \midrule
	2&	8  &	35&	45 &	115\\
        2&	16 &	478&	491&	577\\
	2&	32 &	7422&	7456&	7698\\
	4&	8  &	17&	23&	102\\
	4&	16 &	244&	254&	336\\
	4&	32 &	3753&	3789&	4019 \\
	8&	8  &	8	&15&	95\\
	8&	16 &	121&	136&	216\\
	8&	32 &	1891&	1932&	2053 \\
\bottomrule
    \end{tabular}
    \caption{RPA calculation timings for different parts of the code: the innermost $\bm{q},\omega$ loop, the total RPA routine time, and the one of the total calculation including the DFT reference. Results have been evaluated on hBN for different regular $Q_{G}/K_{G}$ sizes.}
    \label{tab:hbnabstimes}
\end{table}

As a second test, we show in Fig.~\ref{fig:hBN} the scaling for different sizes of $Q_{G}$ and $K_{G}$ for h-BN using regular $\bm{k}$-grids. Absolute timings are shown in table~\ref{tab:hbnabstimes}.
Here, the scaling is again very close to the theoretical limit.
The deviation of the $8 \times 8$ cases showing a super-perfect scaling likely derives from an artefact in the reference $2 \times 192 $ cores case, normalised to $100\%$. This is due to some operations that are independent of the $\bm{k}$-grid size and therefore are more relevant for the smallest grid. These deviations are practically negligible and do not affect our conclusions.

\subsection{Adsorption CO on MgO}
As a practical application of our algorithm, we calculate the adsorption energy of CO on a MgO(001) surface at the RPA@PBE level of theory.
Also known as the 'hydrogen molecule of surface science \cite {Sauer2019}, the adsorption of CO on MgO(001) is one of the most widely studied molecule-surface interactions.\cite{Ugliengo2002, Valero2008, Valero2010, Boese2013, Alessio2019, Mitra2022, Shi2023a, Ye2024, Huang2025} While MgO is relevant as a catalyst for a wide variety of applications,\cite{Smart1973, Ito1985, Morales1989, Ueda1990, Kumar2007} the adsorption of a small molecule on metal oxide surfaces is prototypical for many problems in heterogeneous catalysis and surface science.\cite{Li2016ads, Gui2022ads, Campbell2013ads} Also RPA@PBE results have been reported previously, both using a fully periodic implementation\cite{Bajdich2015} as well as in a finite cluster approach.\cite{Mazheika2016} The RPA@PBE adsorption energy reported by \citet{Bajdich2015} is with -1.65 kcal mol$^{-1}$ significantly higher (less negative) than experimental and recent embedded CCSD(T) results\cite{Shi2023a, Ye2024, Huang2025} which all agree on an adsorption energy of around -4.5 kcal mol$^{-1}$. 

\begin{figure}[hbt!]
    \centering
    \includegraphics[width=0.5\linewidth]{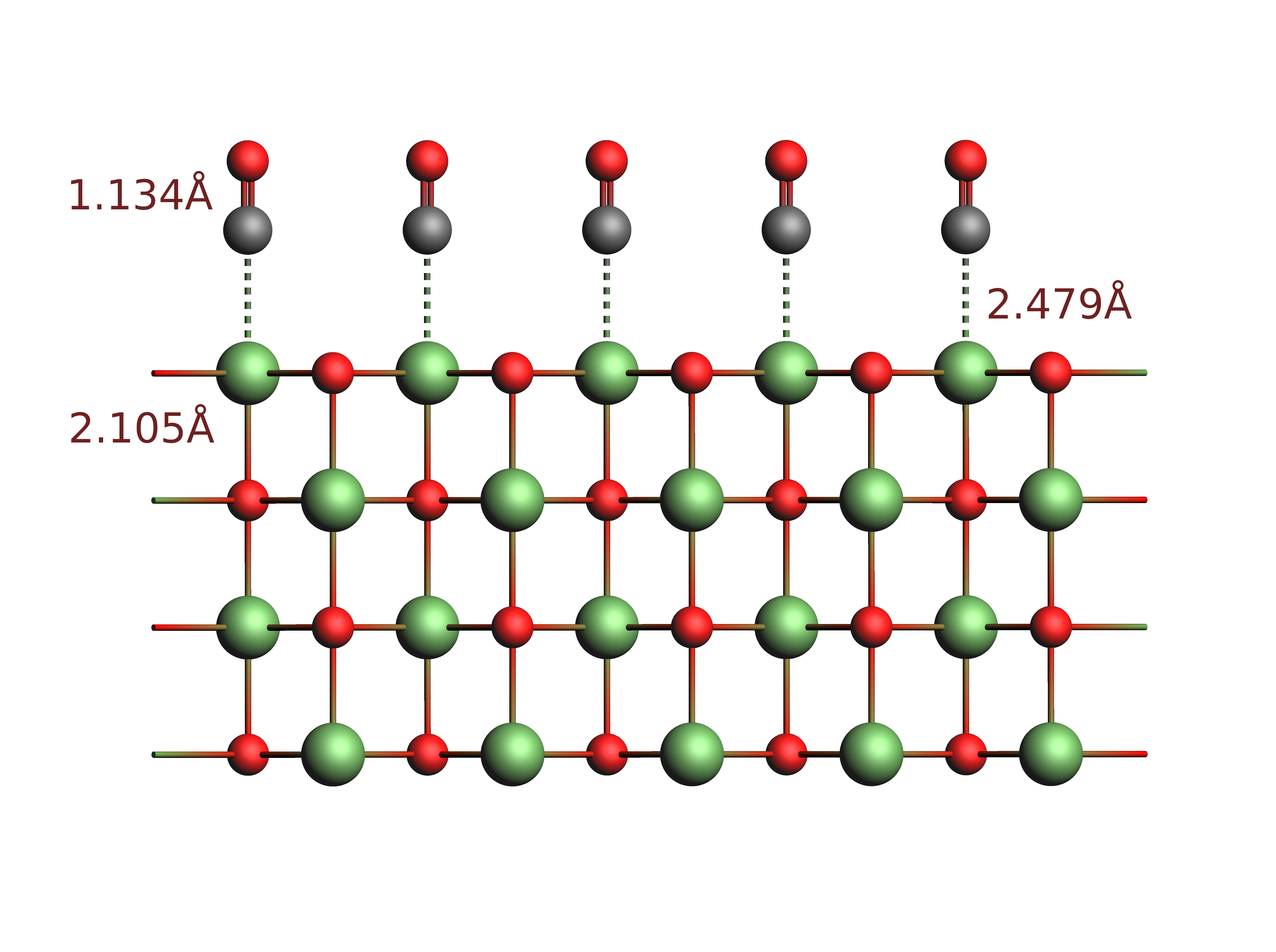}
    \caption{Geometry of Carbon Oxide on MgO adsorption for 100\% coverage case}
    \label{fig:COMGO}
\end{figure}

Here, we focus on the convergence with respect to $K_{g}$ and $Q_{G}$ sampling, basis set convergence, the convergence with respect to the coverage $\theta$, and convergence with the number of MgO layers. As commonly done in the calculation of adsorption energies,\cite{Huang2025, Shi2025} all calculations are counterpoise-corrected to account for basis set superposition errors (BSSE), as this is crucial to obtain reliable adsorption energies.  We do so by using the full basis of the dimer also when computing the energy for the non-interacting systems, yielding:
\begin{equation}
    E^{ads}_c= E^{\text{full basis}}_c(\text{MgO+CO}) - E^{\text{full basis}}_c(\text{CO}) - E^{full basis}_c(\text{MgO}) \;.
\end{equation}
Analysing the BSSE explicitly via the evaluation of
\begin{equation}
    \text{BSSE} = E^{\text{full basis}}_c(\text{MgO}) + E^{\text{full basis}}_c(\text{CO}) - E_c(\text{MgO}) - E_c(\text{CO}) \;,
\end{equation}
is, however, complicated. Using the projector method, the number of functions projected out from the primary basis set of each subsystem does depend on the basis functions present on other subsystems. This gives an additional effect beyond the traditional definition of BSSE,\cite{Boys1970} which, for larger basis sets, where more functions are projected out, can be rather pronounced. 

The system we study is shown in Fig.~\ref{fig:COMGO}. It involves a slab composed of a variable number of MgO layers, and another slab consisting of a single layer of CO molecules of variable density. The MgO surface is modelled through a rock-salt lattice with a lattice constant of 2.105 $\si{\angstrom}$, and the distance between the C and O atoms in the CO molecule is 1.134 $\si{\angstrom}$. The distance between the surface and the CO molecule is the experimental distance of 2.479 $\si{\angstrom}$, and the molecule is oriented perpendicular to the slab \cite{Mitra2022}. All calculations employ dual grids, with an enhancement factor of 0.1. The $K_{g}$ grid consists of a regular number of $\bm{k}$-points, and the total number of $\bm{k}$-points is $5$ times larger than the size of $K_{G}$.

\subsubsection{Convergence with the size of primary and auxiliary basis}

Using PADF, the auxiliary basis used to expand the AO products in \cref{eq:dropPair0} significantly affects the accuracy of correlation energies, especially for larger molecules and basis sets,\cite{Spadetto2023} and their convergence with respect to this parameter is decisive for reliable results. In table~\ref{tab:fitConvergence}, we show the RPA contribution to the counterpoise-corrected adsorption energy of CO on a 2-layer MgO slab at 100 \% coverage for the DZP, TZ2P, and QZ4P basis sets for auxiliary fit sets of variable size. For this test, we employ a 9x9 $\bm{k}$-grid. The threshold parameter for the projector method \cref{projector_threshold} is $\epsilon_{d} = 10^{-3}$. 

\begin{table}[hbt!]
    \centering
        \begin{tabular}{llllcccc}
        \toprule
             &&$N_{bas}$ & $N_{aux}$ & $E_{c}(MgO-CO)$ & $E_{c}(CO)$  & $E_{c}(MgO)$ & $\Delta E_c$ \\
        \midrule
      DZP & 100 & T1 & 873  & -1033.20 & -316.23 & -710.36 & -6.61 \\ 
          &     & T2 & 1477 & -1033.21 & -316.23 & -710.38 & -6.60 \\
                  \midrule
     TZ2P & 172 & T1 & 873 & -1219.42 & -374.14	    & -837.54 & -7.75 \\
          &     & T2 & 1477 & -1218.62 & -374.07	& -836.87 & -7.67 \\
          &     & T3 & 2956 & -1217.88 & -374.06	& -836.29 & -7.53 \\
          &     & T4 & 4070 & -1220.08 & -374.19	& -838.31 & -7.53 \\
                  \midrule
     QZ4P & 273 & T1 & 873 & -1439.00 & -439.05 & -988.74 & -11.22 \\
          &     & T2 & 1477 & -1431.92 & -438.39 & -983.11 & -10.42 \\
          &     & T3 & 2956 & -1419.13 & -437.99 & -972.57 &  -8.56 \\
          &     & T4 & 4070 & -1418.40 & -437.98 & -972.01 &  -8.36 \\
         \bottomrule
    \end{tabular}
    \caption{Total RPA@PBE correlation energies per unit cell and contributions to Adsorption energy in kcal mol$^{-1}$ for 2 MgO layers and 100 \% coverage for different basis sets and auxiliary basis sets using a 9x9 $\bm{k}$-grid. $N_{bas}$ denotes the number of primary basis functions remaining after applying the projector method.}
    \label{tab:fitConvergence}
\end{table}

Different auxiliary basis sets are denoted as T1 to T4. Their generation has been described in Ref.~\citenum{Forster2020}. Importantly, they are not automatically generated from the primary basis. T1 contains auxiliary basis functions up to $l_{max} = 4$, T2 and T3 contain auxiliary basis functions up to $l_{max} = 6$ (As also reflected by the numbers of auxiliary basis functions in each system shown in Table~\ref{tab:fitConvergence}, T3 is composed of a much larger number of fit functions for each angular momentum), and T4 contains additional basis functions with $l_{max} = 7$. Even though the maximum angular momentum in our primary basis sets is $l_{max} = 3$, angular momenta larger than $2l_{max}$ in the auxiliary basis can be important.\cite{Ihrig2015, Spadetto2023} For the DZP basis set, even absolute correlation energies are already converged with the T1 auxiliary basis. The TZ2P calculations show a larger dependence on the auxiliary basis, and the T3 set is necessary for the convergence of the correlation energy difference. For the QZ4P basis set, correlation energies are not converged with T1 and T2, and even the T4 auxiliary basis changes the relative correlation energy by 0.2 kcal/mol compared to T3. The following DZP and TZ2P calculations are all performed with the T2 fit set, and the TZ2P results are adjusted for the resulting fit error of 0.14 kcal/mol in the RPA contribution to the adsorption energy.

\begin{table}[hbt!]
    \centering
        \begin{tabular}{llcccc}
        \toprule
    $\epsilon_{d}$ &$N_{mod}$ & $E_{c}(MgO-CO)$ & $E_{c}(CO)$  & $E_{c}(MgO)$ & $\Delta E_c$ \\
        \midrule
    $1e^{-3}$ & 12 & -1217.88 & -374.06 & -836.29 & -7.53 \\
    $5e^{-4}$ &  8 & -1227.49 & -375.01 & -844.97 & -7.51 \\
    $1e^{-4}$ &  4 & -1234.80 & -377.40 & -850.05 & -7.35 \\
         \bottomrule
    \end{tabular}
    \caption{Total RPA@PBE correlation energies per unit cell and contributions to Adsorption energy in kcal mol$^{-1}$ for 2 MgO layers and 100 \% coverage calculated with the TZ2P basis set and T3 auxiliary basis set. $N_{mod}$ denotes the number of primary basis functions that have been projected out, and $\epsilon_{d}$ is the value of the threshold for the projector method.}
    \label{tab:depThreshold}
\end{table}

After assessing the influence of the auxiliary basis, we also quantify the impact of the threshold for the projector method. We do this here for the TZ2P basis set and the T3 auxiliary basis. The results in Table~\ref{tab:depThreshold} reveal a pronounced sensitivity of absolute RPA correlation energies to this value. This is expected, since a smaller number of $N_{mod}$ translates into a larger primary basis. The contribution to the adsorption energy is relatively stable with respect to this parameter, but increases for $\epsilon_d = 1e^{-4}$. This can be fixed using a larger auxiliary basis set. Repeating the same calculation with the T4 auxiliary basis set again reduces $\Delta E_c$ to -7.50 kcal/mol. This demonstrates that the convergence of a calculation with respect to the size of the auxiliary basis is heavily influenced by the threshold used for the projector method. In all of the following calculations, we will use $\epsilon_{d} = 1e^{-3}$.

\begin{figure}
    \centering
    \includegraphics[width=0.8\linewidth]{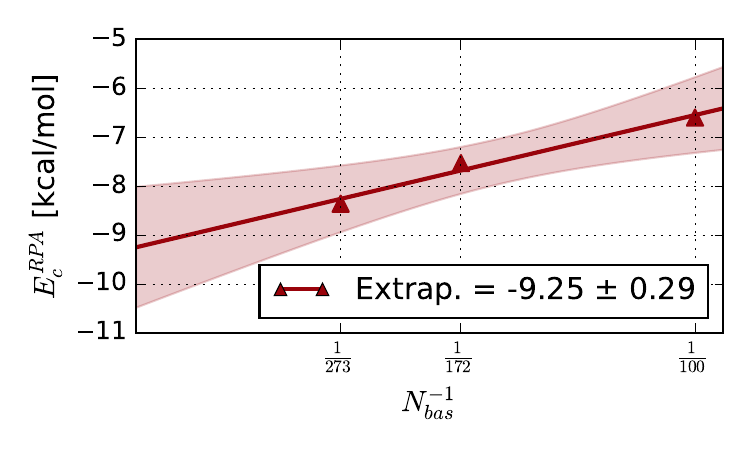}
    \caption{Inverse linear fit of the RPA contribution to the adsorption energy in kcal/mol of CO on a 2-layer MgO slab against the size of the single particle basis. The red area is the confidence interval of the fit.}
    \label{fig:bas_extra}
\end{figure}

Using the numbers that are converged with respect to the auxiliary basis set, we extrapolate the RPA contribution to the adsorption energy to the complete basis set (CBS) limit. We assume a linear dependence of the basis set error on the inverse number of basis functions, as often done in plane wave calculations.\cite{Harl2008} As shown in Fig.~\ref{fig:bas_extra}, the linear fit including all three basis sets (DTQ) captures the trend in the calculated values quite well. From this fit, we obtain an extrapolated RPA contribution to the adsorption energy of –9.25 $\pm$ 0.29 kcal/mol, which is about 1 kcal/mol lower than the QZ result. We assume the standard fit error (intercept) of 0.29 kcal/mol to represent the uncertainty of our extrapolation.

\subsection{$K_{G}/Q_{G}$ convergence}

\begin{figure}[hbt!]
    \centering
    \includegraphics[width=0.8\linewidth]{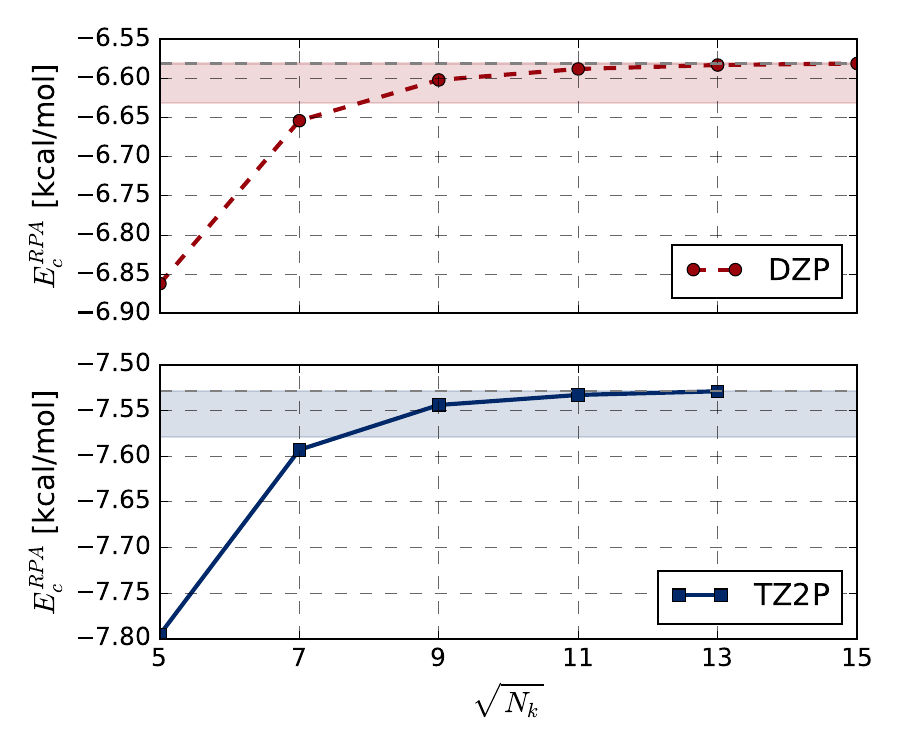}
    \caption{RPA@PBE contribution to adsorption energy in kcal mol$^{-1}$ for 2 MgO layers and 100\% coverage for the DZP and TZ2P basis sets with different $\bm{k}$-grids. The shaded areas highlight the regions where the correlation energy is converged within 0.05 kcal/mol.}
    \label{fig:2l_kconverge}
\end{figure}

Figure~\ref{fig:2l_kconverge} shows the convergence with respect to the number of $\bm{k}$-points of the RPA contribution to the adsorption energy of CO on a 2-layer MgO slab at 100 \% coverage for the DZP and TZ2P basis sets. As for the previous calculations involving the monolayer MgO slab, the convergence with the size of the $\bm{k}$-grid is rapid, and independent of the employed basis set. For both basis sets, convergence within 0.02 kcal/mol is reached already for the $9 \times 9$ grid. In practical calculations, the independence of the convergence rate with respect to the thermodynamic limit is particularly useful, as it allows for converging the RPA correlation energies with respect to the $\bm{k}$-grid in a small basis set, and subsequently to converge them with respect to the single-particle basis size using a coarser $\bm{k}$-grid.

\subsubsection{Number of layers}

Next, we evaluate the RPA contribution to the adsorption energy as a function of the number of MgO layers, using both the DZP basis set. As shown in Table~S13 in the supporting information, this contribution consistently remains at –6.60 kcal/mol, regardless of the number of layers, with the same behavior observed for the TZ2P basis set. This is in agreement with previous correlated calculations for the same system,\cite{Ye2024} where convergence of the MP2 contribution to the adsorption energy has been obtained already with 2 MgO layers as well. 

\subsubsection{Convergence with respect to coverage}

\begin{table}[hbt!]
    \centering
    \begin{tabular}{lcccc}
        \toprule
          & \multicolumn{4}{c}{$\sqrt[dim]{n_{k}}$}   \\
          Coverage & 5&7&   9 & 11  \\ 
          \midrule
          100   & -1.10 &-0.90 & -0.85 & -0.84 \\
           50    & &	&  0.33 &	 0.32 \\
           25$^*$  & \textit{0.07} &	\textit{0.08}  && \\
         \bottomrule
    \end{tabular}
    \caption{RPA@PBE interaction energies of CO with a 2-layer MgO slab for different $\bm{k}$ meshes and coverages in kcal/mol using TZ2P and DZP basis sets. The TZ2P result for 25 \% coverage has been extrapolated by adding the difference between the TZ2P and DZP result for 50 \% and 25 \% coverage to the TZ2P result at 25 \% coverage, as shown in the supporting information in section S.3.2.2.}
    \label{tab:mgococov}
\end{table}

Finally, we investigate the convergence of the adsorption energy with respect to the coverage of the Mg atoms, using a slab geometry with two layers. To this end, we calculate the CO-MgO adsorption energies for the cases of 100 \%, 50 \%, and 25 \% coverage, and show the results in table ~\ref{tab:mgococov}. We obtain an adsorption energy of -0.84 kcal mol$^{-1}$ at 100 \% coverage. In the supporting information (S.3.2.2), we also show that the differences between the 100 \% and 50 \% coverage cases are exactly the same for the DZP and TZ2P basis sets, from which we conclude that the change of the adsorption energy with respect to the coverage is basis set independent. We therefore apply a correction evaluated with the DZP basis set to the TZ2P result at 50 \% coverage to extrapolate it to the 25 \% coverage case. The adsorption energies for the 25 \% and 50 \% cases are relatively close, suggesting that the convergence with respect to the CO saturation is achieved already at the 50 \% case. For this reason, we decided to use the average between the interaction energies obtained for the 50 \% and 25 \% coverages as the error bar for the low coverage limit, amounting to
\begin{align}
    \epsilon_{cov} = \frac{E_{50\%} - E_{25\%}}{2} = 0.12 \text{ kcal mol}^{-1} \;,
\end{align}
while the low-coverage limit is assumed to be reached at 25 \% coverage, obtaining the coverage correction as
\begin{align}
    \Delta_{cov} = E^{2l}_{100\%} - E^{2l}_{25\%} = 0.92 \text{ kcal mol}^{-1}
    \label{eq:covshift}
\end{align}

\subsubsection{Final estimate and discussion}
To obtain the final adsorption energy, we combine the CBS-limit extrapolated RPA contribution from Fig.~\ref{fig:bas_extra} with the PBE and the HF@PBE contributions. We calculated these numbers using 4 MgO layers and the QZ4P basis set, which ensures convergence with respect to both parameters, and we obtain (See Table S14 for the raw data)
\begin{align}
    E^{100\%}_{ads} & = E^{c^{\text{CBS}}}_{RPA} + E^{\text{QZ4P}}_{HF} + E_{PBE}^{\text{QZ4P}} - E^{\text{QZ4P}}_{XC@PBE} \\ 
    & \nonumber = -9.25 \text{ kcal mol}^{-1} - 3.01 \text{ kcal mol}^{-1} - 3.49  \text{ kcal mol}^{-1} + 13.16  \text{ kcal mol}^{-1}\\
    & = -2.59\text{ kcal mol}^{-1} \nonumber \;.
    \label{eq:ads100}
\end{align}
The only source of error in this number stems from the basis set extrapolation, which we estimated above to be 0.29 kcal/mol. To give the final estimate of the MgO(001)-CO adsorption energy, we use as a reference the RPA@PBE/BSSE adsorption energy at 100\% coverage. As mentioned, we apply the constant shift evaluated in ~\cref{eq:covshift} to correct for the low coverage limit. 
Since we consider errors arising from the number of layers and the reciprocal space convergence negligible, we combine the uncertainties from the extrapolation to the CBS error and of the coverage correction to estimate the total uncertainty as
$ e= \sqrt{\epsilon_{bas}^{2} + \epsilon_{cov}^{2}} =0.31 \text{ kcal mol}^{-1}$. The final estimate for the RPA@PBE adsorption energy of CO on the MgO surface in the low coverage limit becomes
\begin{equation}
    E^{\text{CO-MgO}}_{ads}  \approx E^{100\%}_{ads} + \Delta_{cov} \pm e 
     = −1.67 \pm 0.31 \text{ kcal mol}^{-1} \;.
\end{equation}
This result is in excellent agreement with the periodic RPA@PBE calculations by \citet{Bajdich2015} who obtained an adsorption energy of -1.64 kcal mol$^{-1}$ for the same system. Our calculations confirm that RPA@PBE underestimates both the experimental adsorption energy of -4.57 kcal mol$^{-1}$,\cite{Shi2023a} and recent (embedded) CCSD(T) results obtained by different authors.\cite{Shi2023a, Ye2024, Huang2025} This is expected since RPA@PBE is known to underestimate the magnitude of non-covalent binding energies.\cite{Ren2012a, Nguyen2020, Forster2022a}. We assume that better agreement with these values can be reached by evaluating the RPA adsorption energy with orbitals obtained from a hybrid DFT calculation. Using a hybrid scheme\cite{Ren2011} where the RPA correlation energy is obtained with exact Fock exchange from a self-consistent HF calculation, \citet{Bajdich2015} obtained an adsorption energy of -7.14 kcal mol$^{-1}$. It is further known that RPA@PBE0 gives more negative non-covalent molecular interaction energies than RPA@PBE.\cite{Forster2022a} Both findings suggest that RPA@PBE0 will yield more negative energies for the adsorption of small molecules on transition metal surfaces than RPA@PBE. Such calculations require the fully self-consistent evaluation of periodic HF exchange. We are planning to report the results of such calculations shortly.

\section{\label{sec::conclusions}Conclusions}
In this work, we reported the implementation of the RPA correlation with localised atomic orbitals and pair-atomic density fitting\cite{Spadetto2023} in the BAND module\cite{TeVelde1991} of AMS.\cite{Baerends2025} We have introduced an algorithm, inspired by the staggered mesh method of Lin and coworkers\cite{Xing2021, Xing2022} which improves the treatment of the $\Gamma$-point divergence. This allowed us to achieve fast and reliable convergence to the infinite $\bm{k}$-grid limit. A complete account of this algorithm and a detailed benchmarks will be presented in future work. By distributing the evaluation of the most computationally involved steps of the RPA energy evaluation over multiple ScaLapack contexts for all possible pairs of $\bm{q}$ and $\omega$, our code achieves almost perfect parallel efficiency.

We have demonstrated the efficacy of our implementation through an application to the adsorption of CO on MgO(001). After careful extrapolation to the complete basis set and infinite $\bm{k}$-grid limit we obtain a final adsorption energy of $−1.67 \pm 0.31 \text{ kcal mol}^{-1}$ in the low-coverage limit, in excellent agreement with a previous calculation.\cite{Bajdich2015} As could be expected,\cite{Ren2012a, Nguyen2020, Forster2022a} this value underestimates both the experimental adsorption energy of -~4.57 kcal mol$^{-1}$,\cite{Shi2023a} and recent (embedded) CCSD(T) results.\cite{Shi2023a, Ye2024, Huang2025} The remaining observed discrepancies could possibly be addressed by evaluating the RPA correlation energies with orbitals obtained from a hybrid DFT calculation.\cite{Forster2022a} Another promising option to improve over the RPA would be to extend our current algorithm to $\sigma$-functionals,\cite{Trushin2021, Fauser2021, Erhard2022} which can be achieved through minor modifications of the RPA algorithm at the same computational cost.\cite{Fauser2024} Also renormalised adiabatic xc kernel methods have been shown to provide more accurate results than the RPA at comparable computational cost.\cite{Olsen2012a, Olsen2014, Olsen2019}

The developments presented here lay the groundwork for further extensions of our algorithm. In particular, we currently focus on extending our algorithm to metallic systems, which would enable the calculation of accurate energy profiles for the dissociative chemisorption of small molecules on transition metal surfaces. This process is decisive in heterogenous catalysis,\cite{Kroes2021} but difficult to model with (semi-)local density functionals\cite{Gerrits2020} or even advanced embedded wave function methods.\cite{Yin2018, Zhao2020} On the other hand, the RPA has recently been shown to give accurate reaction barrier heights within chemical accuracy for the challenging cases\cite{Gerrits2020} of H$_2$+Cu(111)\cite{Wei2023, Oudot2024} and H$_2$+Al(110).\cite{Oudot2024} The widespread availability of efficient RPA implementations with localised AOs could pave the way for more applications of the RPA to such systems in the future.\cite{Sauer2024, Kroes2025} Another important application would be the calculation of potential energy surfaces for the adsorption of graphene on metal surfaces, which allows for the tuning of its band gap\cite{Olsen2013} with potential applications in catalysis.\cite{Pykal2016} The competition between chemi- and physisorption induced by the interplay of covalent and dispersive interactions between both surfaces is well described by the RPA\cite{Olsen2011, Mittendorfer2011, Olsen2013} but less so by most van der Waals density functionals.\cite{Hamada2010}

Furthermore, we are currently working on evaluating the RPA polarizability in imaginary time.\cite{Shi2024, Shi2025} Combined with PADF, this should lead to an RPA algorithm that scales cubically in the number of atoms\cite{Forster2020b, Spadetto2023, Shi2024, Shi2025} and only linearly in the number of $\bm{k}$-points.\cite{Shi2024, Shi2025} This could be important for applications in heterogeneous catalysis requiring large unit cells. Together, these advancements should position the RPA as a versatile and accurate tool for computational materials science, offering a pathway for tackling increasingly challenging problems in heterogeneous catalysis.

\appendix

\section{RPA Equations in a basis}
In this appendix, we discuss how the expression for the RPA correlation energy can be expressed in a non-orthogonal auxiliary basis set. Given an operator $A(r,r')$ in a $L^{2}$-Hilbert space, we may write 
\begin{align}
     A(r,r') & = \sum_{\alpha\beta\gamma\delta} S^{-1}_{\alpha\gamma} \hspace{3pt} \langle f_{\gamma}(r) | \hat{A} |f_{\delta} (r) \rangle \hspace{3pt} S^{-1}_{\delta \beta} \hspace{5pt} f_{\alpha}(r)f^{*}_{\beta}(r') \\ 
    & =\sum_{\alpha\beta} A_{\alpha \beta } \hspace{5pt} f_{\alpha}(r)f^{*}_{\beta}(r')
    \label{eq:opExpansion}
\end{align}

Where  we have defined 
\begin{align}
    S_{\alpha \gamma}  = \int_{\mathbb{R}^{3}}  f^{*}_{\alpha}(r)f_{\gamma}(r) dr
\end{align}
and 
\begin{align}
    \tilde{A}_{\alpha \gamma} =\langle f_{\alpha} |\hat{A}| f_{\gamma}\rangle =  \int_{\mathbb{R}^{3}}   f^{*}_{\alpha}(r)A(r,r')f_{\gamma}(r') dr dr' \;.
\end{align}
The set of functions $f_{\alpha} \in F$ is assumed to be a basis of the Hilbert space in which $A(r,r')$ is defined. We then obtain
\begin{align}
    Z(r,r'',\omega) & =\int_{\mathbb{R}^{3}} P^{0}(r,r',\omega)v(r',r'') dr' \\ \nonumber
    & =  \int_{\mathbb{R}^{3}} dr' \sum_{\alpha \gamma}  P^{0}_{\alpha \gamma}(\omega) f_{\alpha}(r) f^{*}_{\gamma}(r')    \sum_{\delta \beta}  v_{\delta \beta} f_{\delta}(r') f^{*}_{\beta}(r'')  \\ \nonumber
    & =  \sum_{\alpha \gamma}  \sum_{\delta \beta}   P^{0}_{\alpha \gamma}(\omega) S_{\gamma \delta}    v_{\delta \beta} f_{\alpha}(r) f^{*}_{\beta}(r'')  \\ \nonumber
    & =  \sum_{\alpha  \beta}  \bigg [ P^{0}(\omega) S    v \bigg ]_{\alpha \beta}  f_{\alpha}(r) f^{*}_{\beta}(r'')  \\ \nonumber
    & =  \sum_{\alpha  \beta}  \bigg [ P^{0}(\omega) S S^{-1} \tilde{v} S^{-1} \bigg ]_{\alpha \beta}  f_{\alpha}(r) f^{*}_{\beta}(r'')  \\ \nonumber
    & =  \sum_{\alpha  \beta}  \bigg [ P^{0}(\omega)  \tilde{v} S^{-1} \bigg ]_{\alpha \beta}  f_{\alpha}(r) f^{*}_{\beta}(r'')  \\ \nonumber
     % & = \int dr'' \sum_{\alpha\gamma}\sum_{\delta\beta} P_{\alpha\gamma} [ S^{-1} (f|f) S^{-1} ]_{\delta\beta} f_{\alpha}(q,r)f^{*}_{\gamma}(q,r'') f_{\delta}(q,r'')f^{*}_{\beta}(q,r') \\ \nonumber 
     % & = \sum_{\alpha\gamma}\sum_{\delta\beta} P_{\alpha\gamma} S_{\gamma\delta} \ [ S^{-1} (f|f) S^{-1} ]_{\delta\beta} f_{\alpha}(q,r) f^{*}_{\beta}(q,r')\\ \nonumber
     % & = \sum_{\alpha \beta} [P (f|f) S^{-1} ]_{\alpha \beta}  f_{\alpha}(q,r) f^{*}_{\beta}(q,r')
\end{align}
for the product of polarizability and Coulomb potential.
The trace becomes
\begin{align}
    \text{Tr}[Z(r,r',\omega)] & = \int_{\mathbb{R}^{3}}\int_{\mathbb{R}^{3}}  dr'dr \sum_{\alpha\beta} Z_{\alpha\beta} f^{*}_{\beta}(r')  f_{\alpha} (r) \delta(r,r') \\ \nonumber
    & =  \int_{\mathbb{R}^{3}}\int_{\mathbb{R}^{3}} dr'dr \sum_{\alpha  \beta}  \bigg [ P^{0}  \tilde{v} S^{-1} \bigg ]_{\alpha \beta}  f^{*}_{\beta}(r')  f_{\alpha} (r) \delta(r,r') \\ \nonumber
    & =   \sum_{\alpha  \beta}  \bigg [ P^{0}  \tilde{v} S^{-1} \bigg ]_{\alpha \beta} S_{\beta \alpha} \\ \nonumber
    & = \text{Tr}[\mathbf{P^{0} \tilde{v}}] \\ \nonumber
    & = \text{Tr}[ \mathbf{\tilde{v}^{\nicefrac{1}{2}} P^{0}  \tilde{v}^{\nicefrac{1}{2}}} ]
    \label{eq:RPAbexp}
\end{align}
We can notice that a similar cancellation of the overlap matrix $S$ also occurs in the power series term,
\begin{align}
\frac{1}{n} \text{Tr}(Z(r,r',\omega)^{n})  = \frac{1}{n} \text{Tr} (  [\mathbf{P^{0} \tilde{v} }]^{n} ) \;.
\end{align}

The RPA equation then becomes
\begin{align}
    E^{RPA}_{corr} = - \frac{1}{2\pi} \int_{0}^{\infty} d\omega \text{Tr} \Bigg[ \sum_{n=1}  \frac{1}{n} [ Z(r,r') ] ^{n}   - Z(r,r') \Bigg ]  \\ \nonumber
    = - \frac{1}{2\pi} \int_{0}^{\infty} d\omega  \Bigg[ \sum_{n=1}  \frac{1}{n} \text{Tr} (  [\mathbf{P^{0} \tilde{v} }]^{n} )   - \text{Tr} \left(\mathbf{P^{0} v}\right) \Bigg ]  \\ \nonumber
    = - \frac{1}{2\pi} \int_{0}^{\infty} d\omega \Bigg[ \log \big( \det(\mathbf{1}-\mathbf{Z}) \big ) - \text{Tr} (\mathbf{Z}) \Bigg ] \\ \nonumber
\end{align}
where the matrix $\mathbf{Z}$, by using the cyclic property of the trace operation is defined as $\mathbf{Z}=\mathbf{\tilde{v}^{\frac{1}{2}}P_{0}\tilde{v}^{\frac{1}{2}}}$.
We remark that the discussion presented here assumes the basis $F$ to be complete when representing both the Coulomb potential and the polarizability. In practice, however, this can not be achieved and an error deriving from the incompleteness will always be present.

% \section{Maximum representable length}
% \label{sec:maxlen}
% To discuss \cref{eq:maxgrid} we can consider a 1D Fourier vector composed only of two frequencies.
% \begin{align}
%     s(R) = A e^{ikR} + Be^{i(k+q)R}    
% \end{align}
% We may ask ourselves when this signal repeats itself. This happens exactly when both the terms start a new period.
% \begin{align}
% \begin{cases}
%     (k+q_{min})R_{max} = 2\pi n \\
%     kR_{max} = 2\pi m 
% \end{cases}
% \end{align}
% \begin{align}
% \begin{cases}
%     q_{min}R_{max}  = 2\pi (n-m)  \\
%     kR_{max} = 2\pi m 
% \end{cases}
% \end{align}
% so $R_{max}$ will be obtained when $n-m = 1$
% \begin{align}
%     R_{max} = \frac{2\pi}{q_{min}}
% \end{align}
% This means that, if the smallest increment available in the $\bm{k}$-grid is $q_{min}$, every function beyond $R_{max}$ will repeat itself.

\begin{acknowledgement}
%%%%%%%%%% funding goes here %%%%%%%%%%%%%%%%%
We acknowledge the use of supercomputer facilities at SURFsara sponsored by NWO Physical Sciences, with financial support from The Netherlands Organization for Scientific Research (NWO).
E.S. acknowledges funding from the European Union’s Horizon 2020
research and innovation program under grant agreement No 956813 (2Exciting), private funding from SCM - Software for Chemistry and Materials, and funding from ChemistryNL under grant agreement CHEMIE.PGT.2024.032. A.F. acknowledges funding through a VENI grant from NWO under grant agreement VI.Veni.232.013.
\end{acknowledgement}
\bibliography{bib.bib,all.bib}

%\subfile{SI}

\end{document}